\documentclass[oldversion]{aa}
\usepackage{graphicx,lscape}
\usepackage{lscape}
\usepackage{natbib}
\usepackage{txfonts}
\usepackage{ulem}
     \def\new#1 {{\bf #1 }}
     \def\cut#1 {\sout{#1} }

\def\folio{\ifnum\pageno=1\nopagenumbers\else\number\pageno\fi}

\def\lax    {\ifmmode{_<\atop^{\sim}}\else{${_<\atop^{\sim}}$}\fi}
\def\gax    {\ifmmode{_>\atop^{\sim}}\else{${_>\atop^{\sim}}$}\fi}
\newbox\grsign      \setbox\grsign=\hbox{$>$} 
\newdimen\grdimen   \grdimen=\ht\grsign
\newbox\simgreatbox \setbox\simgreatbox=\hbox{\raise.5ex\hbox{$>$}\llap
                        {\lower.5ex\hbox{$\sim$}}}\ht1=\grdimen\dp1=0pt
\newbox\simlessbox  \setbox\simlessbox =\hbox{\raise.5ex\hbox{$<$}\llap
                        {\lower.5ex\hbox{$\sim$}}}\ht2=\grdimen\dp2=0pt

\newbox\grsign \setbox\grsign=\hbox{$>$} \newdimen\grdimen \grdimen=\ht\grsign
\newbox\laxbox \newbox\gaxbox
\setbox\gaxbox=\hbox{\raise.5ex\hbox{$>$}\llap
     {\lower.5ex\hbox{$\sim$}}}\ht1=\grdimen\dp1=0pt
\setbox\laxbox=\hbox{\raise.5ex\hbox{$<$}\llap
     {\lower.5ex\hbox{$\sim$}}}\ht2=\grdimen\dp2=0pt
\def\gax{\mathrel{\copy\gaxbox}}
\def\lax{\mathrel{\copy\laxbox}}
\def\boxit#1    {\vbox{\hrule\hbox{\vrule\kern3pt
                  \vbox{\kern3pt#1\kern3pt}\kern3pt\vrule}\hrule}}
\def\h      {\ifmmode{^{\rm h}}\else{$^{\rm h}$}\fi}
\def\m      {\ifmmode{^{\rm m}}\else{$^{\rm m}$}\fi}
\def\s      {\ifmmode{^{\rm s}}\else{$^{\rm s}$}\fi}
\def\decas    {\ifmmode{{\rlap.}{''}}\else{${\rlap.}{''}$}\fi}
\def\mum     {\ifmmode{\mu{\rm m}}\else{$\mu{\rm m}$}\fi}
\def\s      {\ifmmode{^{\rm s}}\else{$^{\rm s}$}\fi}
\def\deg      {\ifmmode{^{\circ}}\else{$^{\circ}$}\fi}
\def\as     {\ifmmode {\rlap.}$\,$''$\,$\! \else ${\rlap.}$\,$''$\,$\!$\fi}
\def\decsec  {\ifmmode {\rlap.}$\,$^{s}$\,$\! \else ${\rlap.}$\,$^{s}$\,$\!$\fi}\def\decs  {\ifmmode {\rlap.}$\,$^{s}$\,$\! \else ${\rlap.}$\,$^{s}$\,$\!$\fi}

\def\kms    {\ifmmode{{\rm km~s}^{-1}}\else{km~s$^{-1}$}\fi}

\def\Mspy   {\ifmmode {M_{\odot} {\rm yr}^{-1}} \else $M_{\odot}$~yr$^{-1}$\fi}
\def\Mdot   {\ifmmode {\dot M} \else $\dot M$\fi}
\def\mhd    {\ifmmode {n_{{\rm H}_2}} \else $n_{{\rm H}_2}$\fi}
\def\mhcd   {\ifmmode {N_{{\rm H}_2}} \else $N_{{\rm H}_2}$\fi}

\def\El      {\ifmmode{E_{\ell}}\else{$E_{\ell}$}\fi}
\def\beam    {\ifmmode{\theta_{\rm B}}\else{$\theta_{\rm B}$}\fi}
\def\mjyb   {\ifmmode {{\rm mJy~beam}^{-1}} \else{mJy~beam$^{-1}$}\fi}
\def\mujyb   {\ifmmode {\mu{\rm Jy~beam}^{-1}} \else{$\mu$Jy~beam$^{-1}$}\fi}
\def\Trot   {\ifmmode{T_{\rm rot}}\else$T_{\rm rot}$\fi}    
    
\def\Teff   {\ifmmode{T_{\rm eff}}\else$T_{\rm eff}$\fi}

\def\ITRS   {\ifmmode{\smallint {\rm T}_{R}^{*}dv}\else{$\smallint 
{\rm T}_{R}^{*}dv$}\fi}
\def\ITRS   {\ifmmode{\smallint {\rm T}_{R}^{*}dv}\else{$\smallint 
{\rm T}_{R}^{*}dv$}\fi}
\def\ITAS   {\ifmmode{\smallint {\rm T}_{A}^{*}dv}\else{$\smallint 
{\rm T}_{A}^{*}dv$}\fi}

\def\lefttitle#1  {\noindent \hangindent=18.0pt \hangafter=1 {#1} \par}
\def\vol#1  {{\bf {#1}{\rm,}\ }}
\font\tenssb=cmssbx10
\font\tenbf=cmbx10
\font\sevenbf=cmbx8
\font\fivebf=cmbx6
\def\unetdemi    {\smallskipamount=6pt plus2pt minus2pt
                  \medskipamount=12pt plus4pt minus4pt
                  \bigskipamount=24pt plus8pt minus8pt
                  \normalbaselineskip=16pt plus0pt minus0pt
                  \normallineskip=2pt
                  \normallineskiplimit=0pt
                  \jot=6pt
                  {\def\smallskip {\vskip\smallskipamount}}
                  {\def\medskip   {\vskip\medskipamount}}
                  {\def\bigskip   {\vskip\bigskipamount}}
                  {\setbox\strutbox=\hbox{\vrule 
                    height17.0pt depth7.0pt width 0pt}}
                  \parskip 12.0pt
                  \normalbaselines}
\def\smallerspace {\smallskipamount=3pt plus0pt minus0pt
                  \medskipamount=6pt plus0pt minus0pt
                  \bigskipamount=10.5pt plus0pt minus0pt
                  \normalbaselineskip=10.5pt plus0pt minus0pt
                  \normallineskip=1pt
                  \normallineskiplimit=0pt
                  \jot=3pt
                  {\def\smallskip {\vskip\smallskipamount}}
                  {\def\medskip   {\vskip\medskipamount}}
                  {\def\bigskip   {\vskip\bigskipamount}}
                  {\setbox\strutbox=\hbox{\vrule 
                    height8.5pt depth3.5pt width 0pt}}
                  \parskip 0pt
                  \normalbaselines}
\def\memospace    {\smallskipamount=4pt plus1pt minus1pt
                  \medskipamount=6pt plus2pt minus2pt
                  \bigskipamount=14pt plus6pt minus6pt
                  \normalbaselineskip=14pt plus0pt minus0pt
                  \normallineskip=1pt
                  \normallineskiplimit=0pt
                  \jot=4pt
                  {\def\smallskip {\vskip\smallskipamount}}
                  {\def\medskip   {\vskip\medskipamount}}
                  {\def\bigskip   {\vskip\bigskipamount}}
                  {\setbox\strutbox=\hbox{\vrule 
                    height17.0pt depth7.0pt width 0pt}}
                  \parskip 2.0pt
                  \normalbaselines}
\def\memowidespace    {\smallskipamount=5pt plus1pt minus1pt
                  \medskipamount=7.5pt plus2pt minus2pt
                  \bigskipamount=17.5pt plus6pt minus6pt
                  \normalbaselineskip=17.0pt plus0pt minus0pt
                  \normallineskip=1.25pt
                  \normallineskiplimit=0pt
                  \jot=5pt
                  {\def\smallskip {\vskip\smallskipamount}}
                  {\def\medskip   {\vskip\medskipamount}}
                  {\def\bigskip   {\vskip\bigskipamount}}
                  {\setbox\strutbox=\hbox{\vrule 
                    height21.25pt depth8.75pt width 0pt}}
                  \parskip 2.5pt
                  \normalbaselines}
\begin{document}

\title{Submillimeter absorption from SH$^+$, a new widespread
      interstellar radical, $^{13}$CH$^+$ and HCl}

\author{K. M. Menten
\inst{1} \and F. Wyrowski \inst{1} \and A. Belloche \inst{1} \and R. G\"usten
\inst{1} \and L. Dedes\inst{1} \and H. S. P. M{\"uller} \inst{2}}

\offprints{K. M. Menten}

\institute{
Max-Planck-Institut f\"ur Radioastronomie,
Auf dem H\"ugel 69, D-53121 Bonn, Germany\\
\email{kmenten, wyrowski, belloche, rguesten, ldedes@mpifr-bonn.mpg.de} 
\and
I. Physikalisches Institut,  Universit{\"a}t zu K{\"o}ln, Germany\\
             Z{\"u}lpicher Str. 77, 50937 K{\"o}ln, Germany\\
             \email{hspm@ph1.uni-koeln.de}
}

\date{Received / Accepted}
\titlerunning{Interstellar SH$^+$, $^{13}$CH$^{+}$ and HCl absorption}

\dedication{This contribution is dedicated to the memory of John M. Brown, pioneer of sub\-mil\-li\-meter spectroscopy of free radicals.}

\authorrunning{Menten et al.}

\abstract {We have used the Atacama Pathfinder Experiment  12 m telescope (APEX)
to carry out an absorption study of submillimeter wavelength rotational ground-state
lines of H$^{35}$Cl, H$^{37}$Cl, $^{13}$CH$^+$, and, for the first time, of the
SH$^+$ radical (sulfoniumylidene or sulfanylium).
We detected the quartet of ground-state hyperfine structure
lines of SH$^+$ near 683 GHz with the CHAMP+ array receiver against the strong continuum
source Sagittarius B2, which is located close to the center of our Galaxy. In
addition to absorption from various kinematic components of Galactic center gas, 
we also see absorption at the radial velocities belonging to intervening spiral arms. 
This demonstrates that SH$^+$ is a ubiquitous component of the diffuse interstellar medium. 
We do not find clear evidence for other SH$^+$ lines we searched for, which is partially 
due to blending with lines from other molecules. 
In addition to SH$^+$, we observed absorption from H$^{35}$Cl, H$^{37}$Cl, and 
$^{13}$CH$^+$. The
observed submillimeter absorption is compared in detail with absorption in 3\,mm 
transitions of H$^{13}$CO$^+$ and $c$-C$_3$H$_2$ and the CO $J = 1 - 0$ 
and $3 - 2$ transitions.
}



\maketitle

\section{\label{intro}Introduction}

\subsection{(Sub)millimeter spectroscopy of diffuse interstellar clouds}

Starting in the late 1930s, studies of molecular absorption toward 
diffuse and translucent clouds have traditionally been in the realm of 
optical and, later, ultraviolet (UV) and infrared (IR) spectroscopy. 
Over the last 15 years,
observations at radio and (sub)mm wavelengths have greatly added to our
knowledge of the chemistry of the diffuse interstellar medium (ISM)
\citep[see, e.g., the review of ][]{SnowMcCall2006}. 
Measurements  toward strong continuum sources are a powerful means to
detect spectral lines that have their lower level in (or near) the rotational
ground state of a molecule and may be the only way to detect a number of
species in low density environments or even at all.
In particular, the large  rotational constants of light hydrides,
resulting in a rotational spectrum starting at submillimeter or
far-infrared wavelengths, together with substantial electric dipole moments,
make collisional excitation negligible compared to radiative decay. This
leaves substantial population only in the lowest energy level,
which can be split in the case of fine structure (fs) or hyperfine structure (hfs).

Two particularly fruitful avenues of research have emerged; First, choosing
extragalactic sources with strong mm- and cm-wavelength continuum emission as background objects, H. Liszt and R. Lucas
have studied the abundances of a variety of di-, tri-, and even polyatomic
molecules toward up to 20 lines of sight with often surprising results
\citep[see, e.g., ][and references therein]{LisztLucas2002}. A second possibility is using as a background the
thermal continuum from star-forming regions, (mostly) free-free emission at cm- and longer
mm-wavelengths and dust emission at shorter wavelengths. 
Particularly interesting lines of sights are those toward regions M and N in the Sagittarius
B2 Giant Molecular Cloud (GMC), which allow sensitive absorption spectroscopy of the ISM
surrounding the Galactic center (GC) near which they are located at a distance of $\sim 8$~kpc
\citep{Reid2009} as well as of  intervening GMCs.
For a long time, absorption between $-120$ \kms\ and Sgr~B2's local velocity range 
($> +50$ \kms) has been observed in the HI 21~cm line \citep[see ][and references 
therein]{GarwoodDickey1989} as well as in (mostly) mm-wavelength lines 
from quite a number of molecules \citep[see ][and references therein]{Linke1981, Nyman1983, Greaves1994}. 
Intervening spiral arm absorption toward other bright inner Galaxy (sub)mm continuum sources 
has been found as well, most prominently toward the central radio/(sub)mm source Sgr A$^\star$
and W49N \citep[see, e.g.,  ][]{Liszt1977, Linke1981, Cox1988, Greaves1994, Plume2004, Polehampton2005a}.

At \textit{sub}millimeter wavelengths, numerous molecules have been detected in 
absorption via their ground-state transitions (until recently mostly) toward Sgr~B2 \citep[for 
summaries see][]{Menten2004, Lis2009}. In particular, the  $1_{10} - 1_{01}$ transition of ortho-H$_2^{16}$O and -H$_2^{18}$O was detected with the Submillimeter Wave Astronomy Satellite \citep{Neufeld2000};  this H$_2^{16}$O line, near 557 GHz, is responsible for much of the Earth atmosphere's shorter-wavelength submillimeter opacity.

In fact,  atmospheric absorption by the rotational spectrum of H$_2$O (and to a lesser extent of O$_2$)  makes the ground-state lines of many light hydride species unaccessible from the ground. Using the  Long Wavelength Spectrometer 
(LWS) aboard the Infrared Space Observatory (ISO), which  covered frequencies $> 1.5~$THz (wavelengths $< 200~\mu$m),
several hydroxyl isotopologues 
\citep{Polehampton2003, Polehampton2005b} were detected, as well as  CH$_2$  \citep{Polehampton2005a}, 
and $para$-H$_2$D$^+$ \citep{Cernicharo2007}. From the excellent high altitude site in the Atacama desert, recently OH$^+$ was found with the APEX telescope  by  \citet{wyrowski2010a}.

Very recently, a new era in submillimeter absorption spectroscopy started with operation of the High Frequency Instrument for the Far Infrared 
\citep[HIFI; see ][]{deGraauw2010} aboard Herschel \citep{Pilbratt2010}. Utilizing the superb sensitivity of Herschel's 3.5 m aperture combined with HIFI's high spectral resolution heterodyne receivers operating in a low thermal background space environment, revolutionized observations of light hydrides.
(Mainly) in the framework of the Guaranteed Time Key Projects \textit{Herschel observations of EXtra-Ordinary Sources (HEXOS)}  (PI: E. A. Bergin) and \textit{PRobing InterStellar Molecules with Absorption line Spectroscopy} (PRISMAS, PI: M. Gerin) observations were made of 
H$_2$O$^+$,
H$_3$O$^+$, 
H$_2$O  and its isotopologues, 
H$_2$Cl$^+$, 
C$_3$,
CH,
and CH$^+$, 
as well as NH and NH$_2$, while NH$^+$ has so far remained undetected \citep{Ossenkopf2010, Schilke2010, Gerin2010, Wyrowski2010b, Neufeld2010a, Lis2010a, Melnick2010, Comito2010, Lis2010b, Phillips2010, Neufeld2010b, Sonnentrucker2010, Mookerjea2010, Qin2010, Falgarone2010, Persson2010}.

High quality optical/UV observations, needed, e.g., for meaningful abundance
determinations, mostly sample relatively nearby clouds since they require
optically bright background stars, typically members of OB associations (with HD
catalogue numbers) within a few kpc \citep[see, e.g. ][and references
therein]{Gredel1997, Stahl2008}.  
In contrast, being much less affected by interstellar extinction, radio/submm and near infrared absorption observations, 
e.g. of the key H$_3^+$ molecular ion \citep{GeballeOka1996}, 
are capable of probing diffuse clouds throughout the Milky Way. 

\subsection{\label{shpintro}Interstellar SH$^+$}
The ionized hydride SH$^+$ (sulfoniumylidene or sulfanylium)
potentially is a  probe of comets, photon-dominated regions (PDRs), shocked interstellar
gas and diffuse molecular clouds \citep{Horani1985} . Toward the latter, optical absorption
measurements have shown for a very long time large abundances of another
ionized molecule, methylidynium, CH$^+$,
the first interstellar molecule, detected by \citet{Dunham1937} and identified
by \citet{DouglasHerzberg1941}. CH$^+$'s abundance is orders of magnitude
higher than predicted by otherwise successful models \citep[see,
e.g.,][]{vanDishoeckBlack1986}. This has been a long-standing puzzle to
astrochemists, who invoked shock wave scenarios to overcome the formation
reaction's endothermicity \citep{Federman1982}.

\citet{Millar1986} and also \citet{Pineau1986} have extended chemical models
by including sulfur chemistry and predicted efficient SH$^+$ formation to occur
in shocked diffuse clouds and that the abundance of SH$^+$ can
approach that of CH$^+$ in moderate velocity magnetic shocks, but is negligible
for a non-magnetic shock. This, in principle, could make the CH$^+$/SH$^+$
abundance ratio a good shock indicator.

\citet{Millar1988} made sensitive searches of an optical wavelength electronic
transition  of SH$^+$ from the $A^3\Pi - X^3\Sigma (0,0)$ band at 3363.49 \AA\
toward the well-studied diffuse cloud in the direction of $\zeta$ Oph, which
shows significant CH$^+$ absorption. They were only able to put an upper limit
on the SH$^+$ column density of $1 \times 10^{13}$~cm$^{-2}$, which is a factor of 3
lower than the measured CH$^+$ column density. \citet{MagnaniSalzer1991}
searched for this line and another one at  3366.08 \AA\ toward a sample of 13 stars with
known diffuse molecular clouds in front of them and were only able to determine
comparable $3\sigma$ upper limits on the SH$^+$ column density between 0.4 and
$1 \times 10^{13}$~cm$^{-2}$ and upper limits on the [SH$^+$/CH$^+$] abundance ratio
between 0.12 and 2. An earlier search by \citet{MagnaniSalzer1989} toward a
sample of diffuse and high-latitude clouds resulted in a factor of a few coarser limits on
the SH$^+$ abundance. High (Galactic) latitude clouds belong to the so-called translucent clouds. 

Translucent clouds, which were recognized as a class by \citet{vanDishoeckBlack1986} 
have physical properties in between those of diffuse and dense dark clouds. 
In particular, they have higher visual extinctions, $A_V$ , (1--a few mag) than 
diffuse clouds ($<1$ mag) and, consequently, a higher ratio of molecular to atomic gas. 

Other astrochemical interest in SH$^+$ derives from the fact that, given a
possible formation route via doubly ionized sulfur (S$^{+2}$), it is predicted
to be a diagnostic of X-ray dominated regions (XDRs) \citep{Abel2008}.
\citet{Staeuber2007} report an observation of the SH$^{+}$ $N = 1-0, J =0-1, F
= 0.5 - 1.5$ line near 346 GHz toward the high-mass young stellar object
(YSO) AFGL 2591, which is surrounded by an XDR. Since this line is blended with
a $^{34}$SO$_2$ line, they were only able to quote an upper limit to the
SH$^+$ line's intensity. Again very recently, Herschel/HIFI observations led to the discovery of emission in the 526 GHz $1_2 - 0_1$ transition of SH$^+$ toward the high-mass protostar W3 IRS 5 \citep{Benz2010}.


In \S\ref{labspec} of this paper we review the laboratory spectroscopy picture
for both SH$^+$ and  $^{13}$CH$^+$. In \S\ref{obs} we describe our
submillimeter observations.  Our results and the analysis procedure used to
derive primary line parameters of the lines are presented in \S\ref{analysis}. 
In \S\ref{discussion} we discuss the derived line parameters in an astrochemical context.


\section{\label{labspec}Rest frequencies}

\subsection{Accurate SH$^+$ frequencies}
\label{SH+}

\begin{figure}[t]
\begin{center}
\includegraphics[angle=-90,width=0.9\hsize]{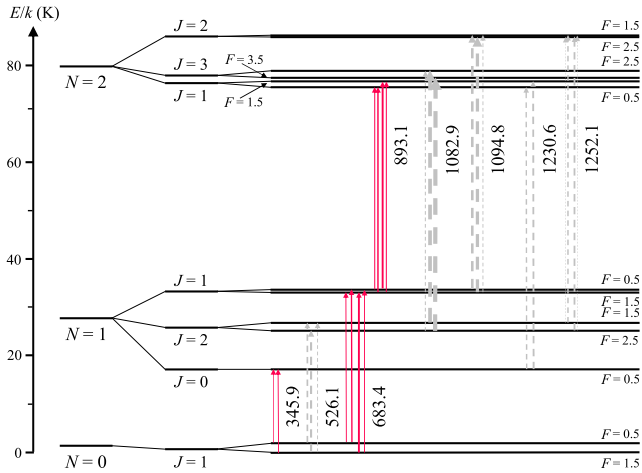}
\caption{\label{levels} {\it Detail of the} energy level diagram of SH$^+$.
The solid arrows mark transitions observed in the course of this work.
The thickness of the arrows indicates the relative strengths of the transitions.
The hyperfine splitting has been exaggerated slightly for clarity reasons.}
\end{center}
\end{figure}

\citet{Savage2004} performed the only measurements of SH$^+$ with microwave accuracy.
They obtained rest frequencies for four hfs lines from the X~$^3\Sigma^-$ $N = 1-0$ state,
one near 346 and three near 526 GHz. A previous study by \citet{Hovde1987}
employed laser magnetic resonance to record rotational spectra.
Recently, \citet{Brown_SH+_2009} used these and further rovibrational (infrared)
data on energy levels of SH$^+$ in and between the $\varv = 0$ and 1 vibrational states
of the $X^3 \Sigma^-$ electronic ground state to perform a single,
weighted least-squares fit in order to determine an improved set of molecular
parameters for this molecule. They used their results to calculate the
rotational spectrum of SH$^+$  in these vibrational levels up to the
$N = 4 - 3$ transition.
The predictions presented in the catalog section\footnote{
http://www.astro.uni-koeln.de/cdms/entries/, see also
http://www.astro.uni-koeln.de/cdms/catalog/} of the Cologne Database for Molecular
Spectroscopy, CDMS, \citep{CDMS1_2001,CDMS2_2005} are based on the data from
\citet{Savage2004}, on weighted averages of the experimental data from
\citet{Hovde1987} extrapolated to zero field taken from the fit of 
\citet{Brown_SH+_2009} and all available infrared data (up to $\varv = 4 - 3$).
These predictions are essentially identical to those in \citet{Brown_SH+_2009}, 
in particular when the uncertainties are taken into account. 
The dipole moment of 1.285~D in the ground vibrational state has been taken from an 
ab initio calculation by \citet{SH+_ai_with_dip_1985}.
The rest frequencies from the CDMS catalog are given in Table~\ref{t:tcont_HSPM3}.
Fig.~\ref{levels} presents the lower part of the rotational energy level diagram
of SH$^+$.

\begin{table*}
 \caption{Spectroscopic data and beam-averaged continuum brightness temperature assumed for the LTE 
 modeling of the absorption lines toward Sgr~B2(M).}
\label{t:tcont_HSPM3}
 \begin{center}
 \begin{tabular}{lccr@{}llrcccl}
  \multicolumn{1}{c}{Transition} & \multicolumn{1}{c}{$F'-F"^a$} & \multicolumn{1}{c}{Position$^b$} & \multicolumn{2}{c}{Frequency} 
& \multicolumn{1}{c}{$S$} & \multicolumn{1}{c}{$E_l$} & \multicolumn{1}{c}{$F_{\mbox{eff}}^c$} & \multicolumn{1}{c}{$B_{\mbox{eff}}^d$} 
& \multicolumn{1}{c}{$T_{\mbox{cont}}^e$} & \multicolumn{1}{c}{reference$^f$} \\
 & & \multicolumn{1}{c}{\scriptsize ($\arcsec,\arcsec$)} & \multicolumn{2}{c}{\scriptsize (MHz)} & 
& \multicolumn{1}{c}{\scriptsize (K)} & & & \multicolumn{1}{c}{\scriptsize (K)} & \\
  \multicolumn{1}{c}{(1)} & \multicolumn{1}{c}{(2)} & \multicolumn{1}{c}{(3)} & \multicolumn{2}{c}{(4)} & \multicolumn{1}{c}{(5)} 
& \multicolumn{1}{c}{(6)} & \multicolumn{1}{c}{(7)} & \multicolumn{1}{c}{(8)} & \multicolumn{1}{c}{(9)} & \multicolumn{1}{c}{(10)} \\
  \hline
$c$-C$_3$H$_2$ 2$_{12}$--1$_{01}$&           & 2.6,--2.0  &  85338&.893(14) & 1.50   & 0.0$^g$& 0.95 & 0.78 & 5.9  & JPL, $h$  \\
  H$^{13}$CO$^+$ 1--0            &           & 2.6,--2.0  &  86754&.2884(46)& 1.0    & 0.0    & 0.95 & 0.78 & 5.9  & CDMS, $i$ \\
  CO 1--0                        &           & 2.6,--2.0  & 115271&.2018(5) & 1.0    & 0.0    & 0.95 & 0.74 & 5.9  & CDMS, $j$ \\
  CO 3--2                        &           & 2.6,--2.0  & 345795&.9899(5) & 3.0    & 16.6   & 0.95 & 0.72 & 11.0 & CDMS, $j$ \\
  SH$^+$ 1$_0$--0$_1$            & $0.5-0.5$ & 0,0        & 345843&.591(128)& 0.2289 & 0.0    & 0.95 & 0.72 & 11.0 & CDMS, $-$ \\
                                 & $0.5-1.5$ &            & 345929&.810(75) & 0.5780 & 0.0    &      &      &      & CDMS, $k$ \\
  H$^{37}$Cl 1--0                & $1.5-1.5$ & 0,0        & 624964&.374(100)& 1.3333 & 0.0    & 0.95 & 0.41 & 33.8 & JPL,  $l$ \\
                                 & $2.5-1.5$ &            & 624977&.821(100)& 2.0    & 0.0    &      &      &      & JPL,  $l$ \\
                                 & $0.5-1.5$ &            & 624988&.334(100)& 0.6667 & 0.0    &      &      &      & JPL,  $l$ \\
  H$^{35}$Cl 1--0                & $1.5-1.5$ & 0,0        & 625901&.603(100)& 1.3333 & 0.0    & 0.95 & 0.41 & 33.8 & JPL,  $l$ \\
                                 & $2.5-1.5$ &            & 625918&.756(100)& 2.0    & 0.0    &      &      &      & JPL,  $l$ \\
                                 & $0.5-1.5$ &            & 625932&.007(100)& 0.6667 & 0.0    &      &      &      & JPL,  $l$ \\
  SH$^+$ 1$_1$--0$_1$            & $1.5-0.5$ & 0,0        & 683336&.08(51)  & 0.1890 & 0.0    & 0.95 & 0.41 & 30.0 & CDMS, $m$ \\
                                 & $0.5-0.5$ &            & 683361&.98(48)  & 0.3778 & 0.0    &      &      &      & CDMS, $-$ \\
                                 & $1.5-1.5$ &            & 683422&.31(50)  & 0.9441 & 0.0    &      &      &      & CDMS, $m$ \\
                                 & $0.5-1.5$ &            & 683448&.20(50)  & 0.1887 & 0.0    &      &      &      & CDMS, $m$ \\
  $^{13}$CH$^+$ 1--0             &           & 0,0        & 830216&.096(22) & 1.0    & 0.0    & 0.95 & 0.35 & 47.7 & CDMS, $n$ \\
  SH$^+$ 2$_1$--1$_1$            & $0.5-0.5$ & 0,0        & 893047&.60(95)  & 0.2889 & 32.8   & 0.95 & 0.35 & 43.5 & CDMS, $-$ \\
                                 & $0.5-1.5$ &            & 893073&.49(96)  & 0.1445 & 32.8   &      &      &      & CDMS, $-$ \\
                                 & $1.5-0.5$ &            & 893107&.92(105) & 0.1444 & 32.8   &      &      &      & CDMS, $-$ \\
                                 & $1.5-1.5$ &            & 893133&.81(94)  & 0.7225 & 32.8   &      &      &      & CDMS, $-$ \\
  \hline
 \end{tabular}
 \end{center}
 Notes:
 $^a$Hyperfine quantum numbers as far as appropriate. 
 $^b$Equatorial offset (J2000) with respect to Sgr~B2(M). (0,0) corresponds to 
 $(\alpha,\delta)_{\rm J2000} = $ 
 $17^{\rm h}47^{\rm m}20^{\rm s}\rlap{.}2, -28^\circ23'05''$.
 $^c$Forward efficiency.
 $^d$Main-beam efficiency.
 $^e$The beam-averaged continuum brightness temperature was derived from the 
depth of the saturated absorption lines in the 3 mm wavelength range, and from 
a direct measurement of the baseline level at higher frequency.
 $^f$Catalog used for rest frequency and reference for line measurement, 
see also section~\ref{labspec}.
 $^g$Lower level is $ortho$ ground state; it is 2.35~K above the zero level.
 $^h$\citet{c-C3H2_Vrtilek_1987}.
 $^i$\citet{Schmid-Burgk2004}.
 $^j$\citet{CO_rot_1997}.
 $^k$\citet{Savage2004}.
 $^l$\citet{HCl_rot_1971}.
 $^m$\citet{Hovde1987,Brown_SH+_2009}.
 $^n$\citet{CH+_Amano}
\end{table*}

\subsection{\label{chplusfrequency}The $J = 1-0$ transition of $^{13}$CH$^+$}

The X~$^1\Sigma^+$ $J=1-0$ rotational ground-state transition of the main isotopologue
of methylidynium, $^{12}$CH$^+$, near 835.2~GHz, cannot be observed from the ground 
for objects with Galactic radial velocities because of a strong telluric O$_2$ line 
approximately 1.1~GHz lower \citep{O2_rot_2003,O2_rot_2010}.  
Observations of the $J = 1-0$ line of $^{13}$CH$^+$ at an $\sim$5~GHz lower frequency 
are however possible. 

\citet{Falgarone2005} reported its detection toward the bright star-forming region 
G10.6$-$0.4 and evaluated from their observation the rest frequency as $830132 \pm 3$~MHz.
Two alternative values for the rest frequency, $830107 \pm 1$ or $830193 \pm 4$~MHz, 
remained viable due to ambiguities in ascribing the observed feature to LSR 
velocity components possibly associated with this source. These were discarded 
on the basis of the poor agreement with the prediction of \citet{PearsonDrouin2006} 
who had reported a transition frequency for $J=1-0$ of $^{12}$CH$^+$ and predicted 
the corresponding $^{13}$CH$^+$ rest frequency as 830134.3~MHz. 
This value was based on a combined analysis of their frequency and 
of optical CH$^+$ data of four isotopologues employing an isotopically scaled
Dunham analysis.

Our observations of $^{13}$CH$^+$, however (see \S\ref{obs}) indicate 
a rest frequency of around 830.2~GHz if we assume that this cation 
has an absorption profile similar to those of SH$^+$, HCl, H$^{13}$CO$^+$, 
or $c$-C$_3$H$_2$.

\citet{CH+_HSPM} performed an analysis similar to the one published by 
\citet{PearsonDrouin2006} in order to evaluate rotational data for 
isotopologues of CH$^+$ independently. Applying appropriate parameters 
which describe the breakdown of the Born-Oppenheimer approximation, 
he obtained a rest frequency for $^{13}$CH$^+$ $J = 1-0$ compatible 
with the present observations. The reviewer of \citet{CH+_HSPM} pointed 
out new laboratory measurements of the $J = 1-0$ transitions of 
$^{12}$CH$^+$, $^{13}$CH$^+$, and $^{12}$CD$^+$ which were submitted 
subsequently by \citet{CH+_Amano}. Not only was the value for $^{13}$CH$^+$ 
in quite good agreement with the estimate from our observations, 
but it was also shown that the previously reported value for $^{12}$CH$^+$ 
was incorrect by several tens of megahertz. 
Including the data from \citet{CH+_Amano} in his analysis, 
\citet{CH+_HSPM} calculated transition frequencies for several isotopologues 
of CH$^+$ which now should be quite reliable and which are in 
the CDMS catalog. $^{13}$CH$^+$ $J = 1-0$ value is again given in 
Table~\ref{t:tcont_HSPM3}.

The dipole moment of 1.683~D for CH$^ +$ in the ground vibrational state has been taken 
from an ab initio calculation by \citet{CH+_ai_with_dip_2007}; the change in dipole 
moment upon substitution of $^{12}$C with $^{13}$C is assumed to be small. 
A value of 1.7~D was used in the CDMS catalog \citep{CDMS1_2001,CDMS2_2005}.

\subsection{HCl, $c$-C$_3$H$_2$, H$^{13}$CO$^+$, and CO}

The rest frequencies for the $J = 1 - 0$ transitions of 
both H$^{35}$Cl and H$^{37}$Cl, which have hfs, 
were taken from the JPL catalog~\footnote{website: http://spec.jpl.nasa.gov/ftp/pub/catalog/catdir.html, see also 
http://spec.jpl.nasa.gov/} \citep{JPL-catalog_1998}. 
These frequencies 
were published by \citet{HCl_rot_1971}. 
The dipole moment of H$^ {35}$Cl for $\varv = 0$, $J = 1$ has been determined as 
1.1086~(3)~D by \citet{HCl_HF_1973}; the dipole moment of H$^ {37}$Cl is expected 
to be only marginally different. 
See Table \ref{t:tcont_HSPM3} for the frequencies of HCl and the other species considered.

The estimated accuracy for each of the three hyperfine components is 100~kHz,
sufficient for the present analysis. It should be mentioned in this context that
improved rest frequencies have been published by \citet{HCl_rot_1998} and \citet{Cazzoli_HCl_2004}
employing Lamb-dip spectroscopy.

The $c$-C$_3$H$_2$ frequency data were also taken from the JPL catalog. The $2_{12} - 1_{01}$
transition was measured by \citet{c-C3H2_Vrtilek_1987}; the reported uncertainty
is smaller than the value used in the JPL catalog.

The H$^{13}$CO$^+$ and CO 
rest frequencies
were taken from the CDMS catalog
\citep{CDMS1_2001,CDMS2_2005}. The rest frequency of the $J = 1 - 0$ transition
of H$^{13}$CO$^+$ is from an astronomical observation by \citet{Schmid-Burgk2004},
both CO transitions were from Lamb-dip measurements by \citet{CO_rot_1997}.

\section{\label{obs}Observations and data reduction}
\subsection{Overview}
We have used the Atacama Pathfinder Experiment (APEX) 12\,m telescope for
observations of several sources in a number of submillimeter wavelength 
transitions of SH$^+$ whose accurate frequencies have recently been determined
(see \S\ref{labspec}); these included the quartet of ground-state hfs lines
near 683 GHz. Our observations, described in \S\ref{apexobs}, 
led to the detection of these lines in absorption against Sgr~B2(M) and (N).
In addition to absorption from gas associated with the  GC, we also find absorption 
at the radial velocities belonging to intervening spiral arms, which demonstrates 
that this species is widespread in the Galactic ISM.

For diverse reasons, our results for other SH$^+$ lines  were not as clear
as those obtained for the 683 GHz line, namely the  $N = 1-0, J=0-1$ lines near 346 GHz, 
which are completely blended with the CO $J=3-2$ line and our observations of the 
$N = 2-1, J=1-1$ lines were inconclusive. 
We made, however, an unambiguous detection of absorption in the $J = 1-0$ transition of 
the $^{13}$CH$^+$ radical near 830~GHz, again from gas in the Sgr~B2 region and 
from spiral arm clouds.
Absorption in that line has previously been reported by \citet{Falgarone2005} 
toward G10.6+0.4, leading to an estimate of its rest frequency that is discussed 
in \S\ref{chplusfrequency}. 

Finally, we took a wide bandwidth spectrum that covers the $J = 1 - 0$ transitions of
both the H$^{35}$Cl and H$^{37}$Cl isotopologues of hydrochloric acid. 
The former had been observed by \citet{Zmuidzinas1995} toward a position 
midway between Sgr~B2(M) and (N). Since their spectrum only covered the LSR 
velocity
range between $-10$ and $+140$~km~s$^{-1}$ it does not contain information 
for the spiral arm feature velocities. While we confirm the Sgr~B2 
velocity absorption, no clear absorption is found toward the latter 
in our more sensitive and wider bandwidth spectra. 
In \S\ref{sss:hcl}, our upper limits for these diffuse lines of sight 
are compared with the tentative ultraviolet detection of HCl absorption 
toward $\zeta$ Oph by \citet{Federman1995}.

\subsection{\label{apexobs}APEX observations and data reduction}
Our  SH$^+$ observations were made on 2008 September 19 and 25 under good to
very good weather conditions with the 12-m Atacama Pathfinder Experiment
telescope, APEX\footnote{This
publication is based on data acquired with the Atacama Pathfinder
Experiment (APEX). APEX is a collaboration between the
Max-Planck-Institut f\"ur Radioastronomie, the European Southern
Observatory, and the Onsala Space Observatory.} \citep{Gusten_etal2006}. The precipitable water vapor (PWV) content
was between 0.3 and 0.7 mm throughout the observations.

The 683 and 893~GHz lines were measured in the lower sideband (LSB) with 
the MPIfR-built CHAMP+ receiver array \citep{Kasemann2006}. 
CHAMP+ consists of two modules of 7 pixels each with one central pixel 
and the others forming a hexagon around it. 
The one module covers the $350~\mu$m atmospheric window and the other the 
$450~\mu$m window. 
Calibration was obtained using the chopper wheel technique, 
considering the different atmospheric opacities in the signal and 
image sidebands of the employed double sideband receivers. 
The image sideband was rejected to the 10~dB level. 
During the observations the single sideband system temperatures 
at 683~GHz were between 1700 and 2600~K. 
To ensure flat spectral baselines, the wobbling secondary was chopped 
with a frequency of 1.5~Hz and a throw of $240''$ about the cross elevation axis. 
The wobbler was operated in symmetric mode, which means that source 
and off position are interchanged between subsequent subscans, which cancels 
any asymmetries in the optical paths. 
Such observations deliver a reliable estimate of the continuum level. 
We found, however, that the quality of the baseline deteriorated 
with increasing continuum flux density.

The radiation was analyzed with a new incarnation of the MPIfR Fast Fourier
Transform spectrometer (FFTS)\citep{Klein_etal2006}, which provided two 1.5~GHz wide 
modules set to 1024 frequency channels each. 
The resultant channel spacing was 1.46~MHz, corresponding to 0.49 and 0.70~km~s$^{-1}$ 
at the highest and lowest frequency we observed (893 and 625~GHz, respectively).   
The FFTS modules were operated in series with an overlap of 300~MHz 
to provide a total coverage of 2.4 ~GHz bandwidth. 
To check the telescope pointing, drift scans were made across the continuum 
of Sgr~B2(N).
Pointing corrections were derived from these measurements. 
The pointing was found to be accurate to within $\approx 3''$, 
acceptable given the FWHM beam size, $\theta_{\rm B}$, which is $9''$ FWHM at 683~GHz.

The SH$^+$ line near 346 GHz was observed with the 345~GHz 
atmospheric window module of the Swedish heterodyne facility instrument 
for the APEX telescope (Vassilev et al. 2008) with system temperatures 
between 320 and 370~K.
Two facility FFTS backends with 1 GHz bandwidth and 8192 channels each
were connected to the receiver with a slight overlap to reach a total
bandwidth of 1.8~GHz. The beam width at the observing frequency is about $18''$. 
Pointing was again checked with continuum scans on Sgr~B2(N).

Observations of the $^{13}{\rm CH}^+$ $J=1-0$ lines near 830~GHz were performed
using the
 CHAMP+ array on 2009, August 11 under extremely good weather conditions
 with a PWV content of 0.14 to 0.2~mm. Two 1.5~GHz FFTS modules were used with
 overlap to reach a total bandwidth of 2.4~GHz.  Prior to the
 integration on Sgr~B2, the telescope was pointed on the continuum
 of Sgr~B2(N) and G10.62$-$0.38. For the line observations again the wobbler
 was used with a throw of $240''$  and a chopping frequency of 1.5 Hz.

Finally, observations of the H$^{35}$Cl and H$^{37}$Cl $J = 1 - 0$ transitions 
were made with CHAMP+'s lower frequency module 
in
2009 June under superb weather conditions.

\section{Results and data analysis}
\label{analysis} During our three  observing campaigns, we have gathered data
that allow unambiguous identification of the SH$^+$ and $^{13}$CH$^+$ molecules
toward Sgr~B2. In addition, we have made observations of the $J=1-0$
ground-state transition of both the H$^{35}$Cl and H$^{37}$Cl isotopologues of
hydrogen chloride.

In absorption spectroscopy, the column density in a transition's 
lower energy
level, $N_{\rm l}$, can be calculated from its observed 
optical depth, $\tau$ 
and line width  $\Delta v$ if its excitation temperature, $T_{\rm ex}$, is known:

\begin{equation}
N_l = {{h}\over{8\pi^2}}{{g_{\rm l}}\over{S\mu^2}}\big[1-e^{-h\nu/kT_{\rm ex}} \big]^{-1}\tau \Delta v,
\end{equation}
\noindent
where $A_{ul}$ the Einstein
A coefficient; $h$ and $k$ are the Planck and Boltzmann constants,
respectively, and $\nu = E_u - E_l$ 
the transition's rest frequency; $E_u$ and $E_l$ are its upper and lower energy levels, 
respectively.

The total column density, $N_{\rm tot}$, is given by
\begin{equation}\label{ntot}
N_{\rm tot} = {{N_l}\over{g_l}}e^{E_l/kT_{\rm rot}}Q(T_{\rm rot}),
\end{equation}
where $Q$ is the partition
function for the rotation temperature, $T_{\rm rot}$. 

In the case of  a line with a single absorption 
component, the optical depth and line width are  usually determined by a Gaussian fit
to the  line profile and $ \tau$ is calculated from the 
line to continuum intensity ratio:
\begin{equation}
\tau = -{\rm ln}\,(1 - |T_{\rm L}|/T_{\rm C})
\end{equation}
where $T_{\rm L}$ and $T_{\rm C}$ are the observed line and 
continuum brightness temperature, respectively.

Things are generally not so simple for the immensely line-rich Sgr~B2 sources, 
for which practically the whole (sub)mm range contains a large number of lines 
from different species in each chosen frequency interval, many of them blended. 
To make things even more complex, many of the simpler species show multiple 
absorption components originating from various spiral arms along the line 
of sight that are blended with each other and with spectral components from lines 
of other species at GC velocities. Lines showing hfs, such as SH$^+$ are yet 
more difficult to analyze. 
To deal with this, we employ, as described by \citet{Belloche2008}, the 
XCLASS\footnote{http://www.astro.uni-koeln.de/projects/schilke/XCLASS} 
program created and developed by Peter Schilke. 
XCLASS is an extension of the CLASS spectral line data reduction program 
that is part of the GILDAS\footnote{http://www.iram.fr/IRAMFR/GILDAS/} 
software package developed by the Institute for Radio Astronomy at Millimeter 
Wavelengths (IRAM). Using XCLASS one may fit all lines in a spectrum 
simultaneously as described in \S\ref{ss:modb2m}.

For each species, XCLASS assumes that the level populations are described by 
a single excitation temperature, which we refer to as the rotation temperature, 
$T_{\rm rot}$. 
Different species can have different rotation temperatures.
Here we divide between two cases: 
First, the critical densities of all the lines in the frequency range of interest 
are lower than the density in the modeled emission region. 
Then we are dealing with local thermodynamic equilibrium (LTE), 
which applies, e.g., for the extremely line rich hot cores associated with Sgr~B2(M)
and (N). In particular in the latter, the complex molecules NH$_2$CH$_2$CN,
C$_2$H$_5$OCHO, and C$_3$H$_7$CN along with the $^{13}$C isotopologues
of vinyl cyanide have recently been identified using 
XCLASS with model temperatures of  $\approx 100$--150 K
\citep{Belloche2008,Belloche2009,13C-VyCN_2008}.
In the second, very low density case, which applies for the absorption lines 
discussed in this paper, $T_{\rm rot}$ equals the temperature of the 
cosmic microwave background radiation, $T_{\rm CMB}$, i.e. 2.728~K.
While this might be a good assumption for clouds far from the strong Sgr~B2 
continuum sources, e.g., in the spiral arms, assuming  $T_{\rm rot}$ = $T_{\rm CMB}$ 
will underestimate
 column densities for molecules 
in the vicinity of Sgr~B2(M) and (N).

\subsection{Modeling of the absorption lines toward Sgr~B2(M)}
\label{ss:modb2m}

\subsubsection{The model for the spiral arm and galactic center LSR  velocity features}
The SH$^+$(1$_1$--0$_1$) spectra obtained with the CHAMP+ array consist of the
superposition of a 
number of absorption components associated with the
dense core Sgr~B2(M) and several molecular clouds distributed along the line of
sight with different systemic velocities \citep[see, e.g.,][]{Greaves1994}. 

The
association of the absorption components'   LSR velocities  with Galactic spiral arms 
or kinematic features associated with the GC located along the line of sight is 
presented in Table \ref{t:isorat}. 
It is essentially based on the velocity
interpretation of H$_2$CO and CS absorption by \citet{WhiteoakGardner1979}
and \citet{Greaves1994}, respectively, as well as the analyses of  \citet{Sofue2006} 
and \citet{Vallee2008}. Although in the GC direction absorption close to 
$V_{lsr} = 0$ could in principle come from any location on the line of sight, 
we favor a physical association with the GC since \citet{GardnerWhiteoak1981} 
found low values of the $^{12}$C$/^{13}$C ratio (20--30) for this velocity range, 
consistent with a Galactic center origin.

\begin{figure}[h!]
\centerline{\resizebox{0.7\hsize}{!}{\includegraphics[angle=270]{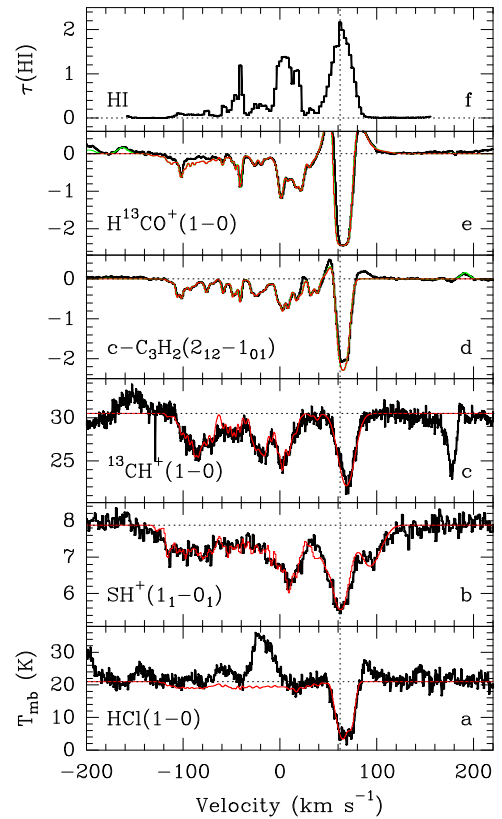}}}
\caption{
\textit{From bottom to top}, panels \textit{a} to \textit{c} show absorption spectra 
taken with the APEX telescope with the CHAMP+ receiver array. Spectra \textit{a} and 
\textit{b} were taken with the central pixel of CHAMP+ toward Sgr~B2(M) 
(see Fig.~\ref{f:shp}).  Spectrum \textit{b} was  obtained toward
a position ($-20.4\arcsec$,$1.4\arcsec$) offset from  Sgr~B2(M) with
the CHAMP+ pixel 2. 
\textit{a}: H$^{35}$Cl $1- 0$ absorption spectrum. 
\textit{b}:  SH$^+$ 1$_1$--0$_1$ absorption spectrum.
\textit{c}: $^{13}$CH$^+$ 1--0 absorption spectrum.
In panels  \textit{a}-- \textit{c}, the red curve
is the synthetic spectrum described in Sect.~\ref{ss:modb2m} and the horizontal
dotted line shows the continuum level. 
\textit{d} and  \textit{e}: same as b for the
$c$-C$_3$H$_2$ 2$_{12}$--1$_{01}$ and H$^{13}$CO$^{+}$ 1--0
transitions observed toward Sgr~B2(M) with the IRAM 30~m telescope,
respectively. Panel \textit{f} shows the optical depth of the 21 cm HI line vs. 
LSR velocity as derived by the procedure described in \S\ref{hicolumn}.
In panels \textit{d} and \textit{e}, the red curve is the synthetic spectrum 
of the molecule, the green curve is the predicted spectrum including in addition 
all molecules identified in our complete 3~mm line survey so far, 
and the horizontal dotted line indicates the zero level (after baseline removal 
because the observations were performed in position-switching mode with a 
far OFF position, which yields a very uncertain baseline level).
In all panels, the spectra are plotted in main-beam brightness temperature scale, 
the vertical dotted line marks the systemic velocity of the source (62~km~s$^{-1}$). 
}
\label{f:3mm}
\end{figure}

The blending of these absorption components is further complicated by 
the hyperfine structure of the SH$^+$ transition. 
To allow meaningful fitting, we used information derived from the absorption spectra 
of other molecules obtained as part of a complete line survey done with the 
IRAM 30~m telescope in the 3~mm atmospheric window toward both hot cores Sgr~B2(M) 
and Sgr~B2(N) \citep[see][]{Belloche2008,Belloche2009}.

The line survey was analysed with the XCLASS software 
(see \S\ref{analysis})\footnote{Where noted, for the components associated 
with Sgr~B2 (around +62 km~s$^{-1}$), a $T_{\rm rot}$ $> T_{\rm CMB}$ was assumed.} . 
Each molecule identified in the survey was modeled with the following free parameters: 
source size, temperature, column density, linewidth, and velocity offset with respect 
to the systemic velocity of the source, 62 km~s$^{-1}$ for Sgr~B2(M). 
When several components were needed to reproduce the observed spectrum of a molecule, 
we fitted a set of parameters for each component. The combination of the predicted 
spectra of all identified molecules was done assuming that the emission adds up 
linearly \citep[for more details, see][]{Belloche2008}.

To model the absorption lines observed toward Sgr~B2(M), we assumed a 
beam-averaged continuum brightness temperature $T_{\rm cont}$ of 5.9~K in
the 3~mm window, as derived from the saturated absorption spectra of HCN and CN 
(see Table~\ref{t:tcont_HSPM3}). The modeling of the absorption components was done 
in the $T_{\rm rot}$ $ = T_{\rm CMB}$  approximation with a flag indicating 
that the absorption spectrum has to be computed against a background emission 
consisting of both the continuum emission and the contribution of the other 
molecules seen in emission, i.e. for each absorption component with the equation

\begin{equation}
T_{\nu} = \left(J_{\nu}(T_{\mathrm{ex}}) - J_{\nu}(T_{\mathrm{cont}})\right) \,\, \left(1 - e^{-\tau_{\nu}}\right) \;\; + \;\; T_{\nu}^{em} \,\, e^{-\tau_{\nu}},
\end{equation}

with $T_{\mathrm{ex}}$ the excitation temperature, 
$T_{\mathrm{cont}}$ the background brightness temperature, 
$T_{\nu}^{em}$ the emission line spectrum, and $\tau_{\nu}$ the opacity 
of the line at frequency $\nu$, for a beam filling factor of 1.

\begin{table}
 \caption{Velocity ranges and $^{12}$C/$^{13}$C isotopic ratios of the Galactic spiral arms along the line of sight of Sgr~B2(M).}
\label{t:isorat}
 \begin{center}
 \begin{tabular}{ccl}
 \hline\hline
 \noalign{\smallskip}
 \multicolumn{1}{c}{Velocity range} & \multicolumn{1}{c}{$^{12}$C/$^{13}$C} & \multicolumn{1}{c}{Location} \\
 \multicolumn{1}{c}{(1)} & \multicolumn{1}{c}{(2)} & \multicolumn{1}{c}{(3)} \\
 \hline \\[-2.0ex]
 $\sim$62     & 20 & Galactic Center, Sgr~B2 \\
    39 --  25 & 40 & Scutum/Crux arm \\
    22 --  12 & 60 & Sagittarius arm \\
    8 --   -9 & 20 & Galactic Center \\
   -13 -- -50 & 40 & 3 kpc expanding ring \\
 $<-53$       & 20 & Galactic Center \\
 \hline

 \end{tabular}
 \end{center}
 \end{table}

That table also lists the $^{12}$C/$^{13}$C isotopologic ratios we assumed.
The $^{12}$C/$^{13}$C isotopic ratio of the Galactic Center components
was fixed to 20 \citep{13C-VyCN_2008}. This value is compatible with
the Galactic gradient derived by \citet[][equation 5]{Milam05}. 
This equation was additionally employed to fix the ratio to 
40 for the Scutum and Norma arm components, and 60 for the 
Sagittarius arm components, with the galactocentric distances of
\citet{Vallee2008}. The $^{16}$O/$^{18}$O isotopic ratio was taken from
\citet{Wilson94}: 250 for the GC components, 327 for the Scutum and Norma arm
components, and 560 for the Sagittarius arm components.


\begin{figure*}[t]
\centerline{\resizebox{0.75\hsize}{!}{\includegraphics[angle=270]{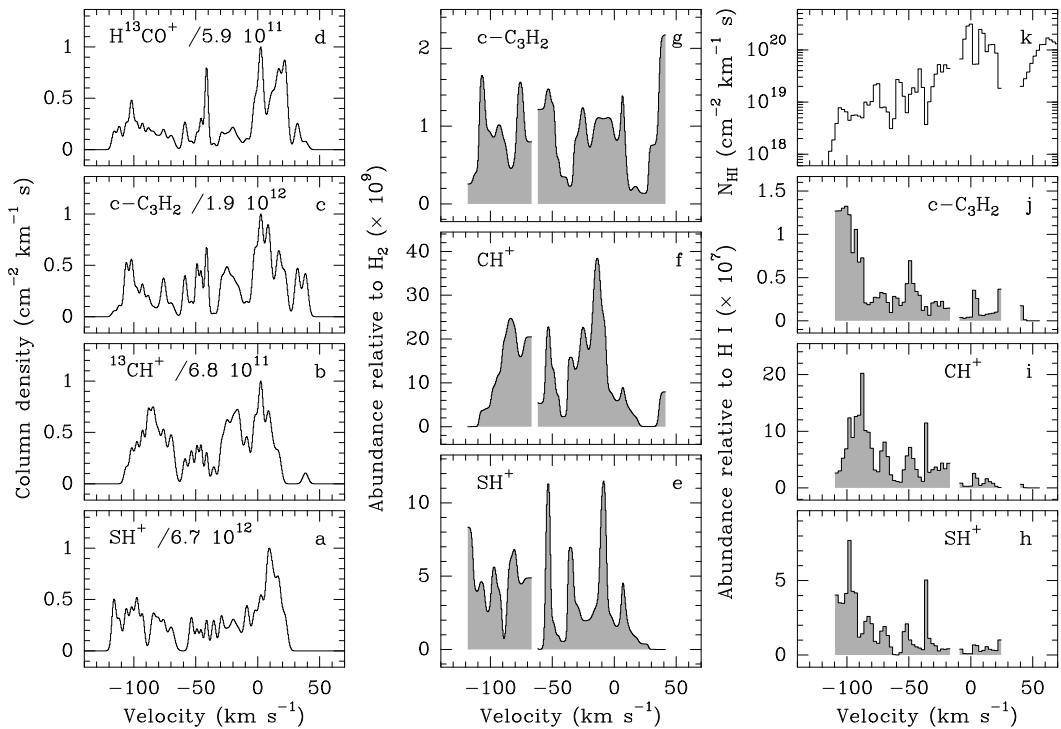}}}
\caption{\textbf{a} to \textbf{d} Column density profiles computed by adding 
the Gaussian velocity absorption  components listed in Table~\ref{t:shpmodel}. 
The components with upper limit were not included.
\textbf{e} to \textbf{g} Abundance profiles relative to H$_2$. The H$_2$ column density 
was computed from the HCO$^+$ column density assuming an HCO$^+$ abundance of 
$5 \times 10^{-9}$. \textbf{h} to \textbf{j} Abundance profiles relative to \ion{H}{i}. 
The \ion{H}{i} column density was computed from comparing an \ion{H}{i} absorption 
with an \ion{H}{i} emission profile (see text).
}
\label{f:coldens}
\end{figure*}

\begin{figure}[t]
\centerline{\resizebox{0.75\hsize}{!}{\includegraphics[angle=270]{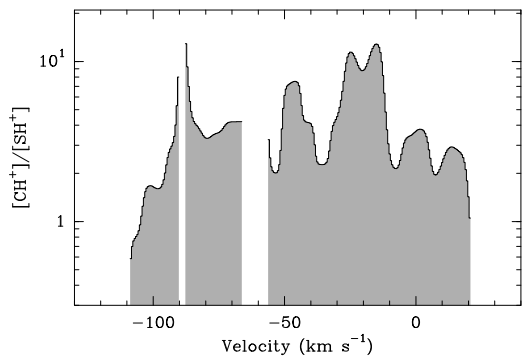}}}
\caption{Ratio of the column density profiles of CH$^+$ and SH$^+$ shown in Fig.~\ref{f:coldens}.
}
\label{f:chpshp}
\end{figure}

We present, in Fig.~\ref{f:3mm} \textit{d} and \textit{e}, the results of our 
modeling for $c$-C$_3$H$_2$(2$_{12}$--1$_{01}$) and H$^{13}$CO$^+$(1--0), respectively. 
The model for H$^{13}$CO$^+$(1--0) was obtained by fitting HCO$^+$, 
H$^{13}$CO$^+$, and HC$^{18}$O$^+$ simultaneously. 
The fit to the observed H$^{13}$CO$^+$(1--0) spectrum consists of four components 
in emission (one for the main emission peak and three to reproduce the wings), 
one absorption component to mimic the self-absorption in 
Sgr~B2(M)\footnote{Although the fits to the self-absorption
features in Fig.~\ref{f:3mm} \textit{d} and \textit{e}  look relatively good, 
we emphasize that a uniform $T_{\rm rot}$ $= T_{\rm CMB}$  model is not complex enough 
to provide reliable physical parameters to describe the self-absorption. 
The parameters listed for the self-absorption components in the following tables 
should therefore be viewed with caution}  
and 34 absorption components associated with the diffuse clouds along the line of sight, 
with linewidths varying between 3 and 8 km~s$^{-1}$ and an excitation temperature 
arbitrarily fixed to 2.7~K.
The parameters of the fit to H$^{13}$CO$^+$ are listed in Table~\ref{t:hcopmodel}.
The model predicts too strong absorption for the
H$^{13}$CO$^+$(1--0) components with $v < -50$ km~s$^{-1}$, while the
corresponding HCO$^+$(1--0) components are well fitted (not shown here). The
disagreement must result either from an underestimate of the $^{12}$C/$^{13}$C
isotopic ratio or from contamination by emission lines that are not yet
included in our complete model. It could in principle also result from larger
excitation temperatures for HCO$^+$ compared to H$^{13}$CO$^+$.

The fit to $c$-C$_3$H$_2$ was done with velocity and linewidth parameters
slightly different from HCO$^+$, although we tried to stay as close as possible
to the components identified in HCO$^+$. The parameters of the fit to
$c$-C$_3$H$_2$ are listed in Table~\ref{t:shpmodel}. The model consists of one
component in emission (we did not attempt to model the possible wing emission
in this case), one absorption component to mimic the self-absorption, and 30
absorption components associated with the diffuse clouds, with an excitation
temperature of 2.7~K and linewidths varying between 3 and 6 km~s$^{-1}$.

\begin{figure*}[t]
\centerline{\resizebox{0.7\hsize}{!}{\includegraphics[angle=270]{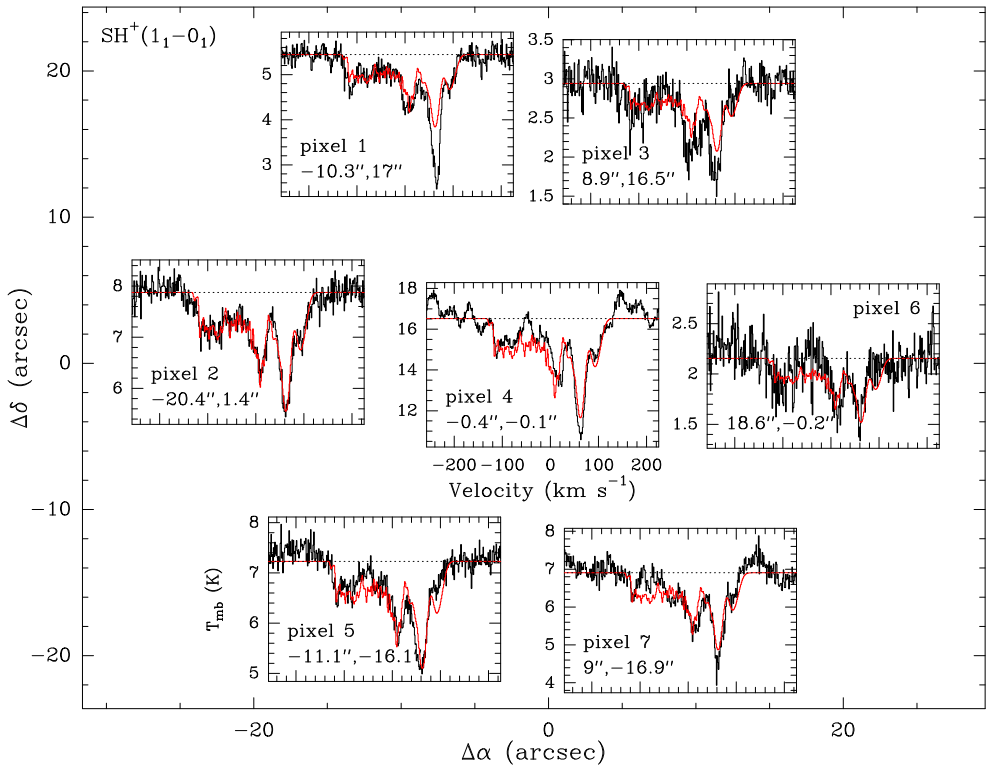}}}
\caption{SH$^+$ absorption spectra obtained toward Sgr~B2(M) with the 7-pixel
array CHAMP+ at the APEX telescope. The equatorial offset position of each
pixel relative to Sgr~B2(M) is given in the bottom left corner of each panel.
In each panel, the dotted line shows the continuum level as measured over the
emission/absorption-free channels, and the red curve is the spectrum predicted
by the $T_{\rm rot}$ $= T_{\rm CMB}$  model described in Sect.~\ref{ss:modb2m}. 
All spectra are plotted in main-beam brightness temperature scale.} \label{f:shp}
\end{figure*}

\subsubsection{Modeling the SH$^+$ absorption}

Since we considered $c$-C$_3$H$_2$  a priori to be a better diffuse gas tracer 
than $^{13}$HCO$^+$ (but see \S\ref{discussion}) and because of the added uncertainty 
of assigning a $^{12}$C/$^{13}$C ratio to each absorption components, 
we decided to model the SH$^+$ absorption spectrum based on the parameters 
derived for $c$-C$_3$H$_2$, i.e. we used the same number of absorption components 
with the same velocities and line widths, except for the component
physically associated with Sgr~B2(M) itself. 
Again, we fixed the excitation temperature of each component to 2.7~K. 
Since the observations were done in wobbler mode, we trusted the baseline level 
and derived the level of continuum emission directly from the observed spectra 
shown in Fig.~\ref{f:shp}. 
To compute the synthetic spectra, we assumed that the physical parameters 
of the absorbing components are uniform across the CHAMP+ field of view 
and therefore we used the same set of parameters for all pixels. 
The resulting synthetic spectra are overlaid on the observed spectra 
in Fig.~\ref{f:shp}. 
The agreement is quite good in general, which gives us some more confidence 
in the way we estimated the continuum level. However, significant deviations are seen 
for pixels 1, 3, and 4. As mentioned in \S \ref{obs}, the baseline quality of the
CHAMP+ array drops with increasing level of continuum emission and line contamination 
from the hot molecular core associated with Sgr~B2(M). 
Pixel 4 is the most affected, and we believe that the poor quality of the fit is 
mainly due to this problem. As far as the northern pixels 1 and 3 are concerned, 
the main absorption component is much  stronger than the one predicted by our 
simple model. This may be due to spatial variations of the SH$^+$ column density
across the Sgr~B2 molecular cloud since the cloud is far from being uniform 
(in particular, the hot, dense core Sgr~B2(N) is located $\sim 50\arcsec$ 
to the north of Sgr~B2(M)).
If the variations from pixel to pixel can be trusted, then we can estimate 
the variations of the SH$^+$ column densities in the spiral arm diffuse clouds 
over the footprint of CHAMP$^+$ by modeling each pixel separately. 
This exercise yields column density variations typically on the order of 
a factor 2, but in some cases up to a factor 5. 
However, given the low signal-to-noise ratio, the unknown baseline quality, 
and the line blending, these numbers are very uncertain and it is unlikely 
that they really reflect the spatial variations of the SH$^+$ column density.

\begin{table*}
 \caption{
 Parameters of the best-fit LTE models of SH$^+$, $^{13}$CH$^+$, and c-C$_3$H$_2$, as well as the alternative LTE model of H$^{13}$CO$^+$ fitted with the same velocity components as c-C$_3$H$_2$.
 }
\label{t:shpmodel}
 \begin{center}
 \begin{tabular}{ccccccccccccccc}
 \hline\hline
  & & & & \multicolumn{1}{c}{SH$^+$} & & \multicolumn{3}{c}{$^{13}$CH$^+$} & & \multicolumn{2}{c}{c-C$_3$H$_2$} & & \multicolumn{2}{c}{H$^{13}$CO$^+$} \\ 
 \cline{5-5} \cline{7-9} \cline{11-12} \cline{14-15}
 \noalign{\smallskip}
  \multicolumn{1}{c}{$n_c^a$} & \multicolumn{1}{c}{$V_{\mathrm{lsr}}$$^{b}$} & \multicolumn{1}{c}{$\Delta V^{c}$} & \multicolumn{1}{c}{$\frac{^{12}\mathrm{C}}{^{13}\mathrm{C}}^d$} & \multicolumn{1}{c}{$N^{e}$} & & \multicolumn{1}{c}{$N^{e}$} & \multicolumn{1}{c}{$R_{13}$$^{f}$} & \multicolumn{1}{c}{$R_{12}$$^{g}$} & & \multicolumn{1}{c}{$N^{e}$} & \multicolumn{1}{c}{$R$$^{h}$} & & \multicolumn{1}{c}{$N^{e}$} & \multicolumn{1}{c}{$R_{12}$$^i$} \\ 
  & \multicolumn{1}{c}{\scriptsize (km~s$^{-1}$)} & \multicolumn{1}{c}{\scriptsize (km~s$^{-1}$)} & & \multicolumn{1}{c}{\scriptsize (cm$^{-2}$)} & & \multicolumn{1}{c}{\scriptsize (cm$^{-2}$)} & & & & \multicolumn{1}{c}{\scriptsize (cm$^{-2}$)} & & & \multicolumn{1}{c}{\scriptsize (cm$^{-2}$)} & \\ 
 \multicolumn{1}{c}{(1)} & \multicolumn{1}{c}{(2)} & \multicolumn{1}{c}{(3)} & \multicolumn{1}{c}{(4)} & \multicolumn{1}{c}{(5)} & & \multicolumn{1}{c}{(6)} & \multicolumn{1}{c}{(7)} & \multicolumn{1}{c}{(8)} & & \multicolumn{1}{c}{(9)} & \multicolumn{1}{c}{(10)} & & \multicolumn{1}{c}{(11)} & \multicolumn{1}{c}{(12)} \\ 
 \hline
 1 & -- & -- & -- & -- & &  -- & -- & -- & & $  2.4 \times 10^{14}$  & -- & &  -- & -- \\  2 & 64.0 & 19.0 &   20 & $  2.0 \times 10^{14}$ & &  $  1.8 \times 10^{13}$ &  10.8 & 0.53 & & $  1.9 \times 10^{14}$  &  1.0 & &  $  1.4 \times 10^{14}$ &   0.07 \\  3 & 38.5 & 5.0 &   40 & $<  2.8 \times 10^{12}$ & &  $  3.8 \times 10^{11}$ &  $<$  7.4 & $<$ 0.18 & & $  4.2 \times 10^{12}$  & $<$  0.7 & &  $  2.4 \times 10^{11}$ &  $<$  0.29 \\  4 & 32.0 & 5.0 &   40 & $<  2.8 \times 10^{12}$ & &  $<  1.6 \times 10^{11}$ & -- & -- & & $  4.7 \times 10^{12}$  & $<$  0.6 & &  $  7.8 \times 10^{11}$ &  $<$  0.09 \\  5 & 22.0 & 5.0 &   60 & $  1.1 \times 10^{13}$ & &  $<  1.6 \times 10^{11}$ & $>$ 70.6 & $>$ 1.15 & & $  4.1 \times 10^{12}$  &  2.8 & &  $  2.5 \times 10^{12}$ &   0.07 \\  6 & 17.0 & 5.0 &   60 & $  2.1 \times 10^{13}$ & &  $  9.6 \times 10^{11}$ &  22.1 & 0.36 & & $  5.9 \times 10^{12}$  &  3.6 & &  $  2.0 \times 10^{12}$ &   0.17 \\  7 & 12.0 & 6.0 &   60 & $  2.8 \times 10^{13}$ & &  $  1.4 \times 10^{12}$ &  19.6 & 0.32 & & $  2.7 \times 10^{12}$  & 10.5 & &  $  1.8 \times 10^{12}$ &   0.25 \\  8 & 8.2 & 5.0 &   20 & $  2.5 \times 10^{13}$ & &  $  2.1 \times 10^{12}$ &  12.2 & 0.60 & & $  8.1 \times 10^{12}$  &  3.2 & &  $  6.0 \times 10^{11}$ &   2.06 \\  9 & 2.5 & 5.0 &   20 & $  1.8 \times 10^{13}$ & &  $  3.5 \times 10^{12}$ &   5.2 & 0.26 & & $  9.7 \times 10^{12}$  &  1.9 & &  $  3.0 \times 10^{12}$ &   0.30 \\  10 & -2.5 & 4.5 &   20 & $  1.1 \times 10^{13}$ & &  $  1.9 \times 10^{12}$ &   5.9 & 0.29 & & $  5.4 \times 10^{12}$  &  2.1 & &  $  1.2 \times 10^{12}$ &   0.46 \\  11 & -8.7 & 5.0 &   20 & $  1.6 \times 10^{13}$ & &  $  1.6 \times 10^{12}$ &   9.7 & 0.48 & & $  1.4 \times 10^{12}$  & 11.1 & &  $  3.2 \times 10^{11}$ &   2.38 \\  12 & -15.5 & 6.0 &   40 & $  8.4 \times 10^{12}$ & &  $  2.9 \times 10^{12}$ &   2.9 & 0.07 & & $  3.2 \times 10^{12}$  &  2.6 & &  $  3.6 \times 10^{11}$ &   0.57 \\  13 & -20.0 & 5.0 &   40 & $  8.4 \times 10^{12}$ & &  $  1.6 \times 10^{12}$ &   5.3 & 0.13 & & $  2.7 \times 10^{12}$  &  3.2 & &  $  5.4 \times 10^{11}$ &   0.38 \\  14 & -25.0 & 6.0 &   40 & $  8.4 \times 10^{12}$ & &  $  2.5 \times 10^{12}$ &   3.3 & 0.08 & & $  5.6 \times 10^{12}$  &  1.5 & &  $  5.4 \times 10^{11}$ &   0.38 \\  15 & -29.5 & 4.0 &   40 & $  8.4 \times 10^{12}$ & &  $  8.0 \times 10^{11}$ &  10.6 & 0.26 & & $  2.1 \times 10^{12}$  &  3.9 & &  $  3.3 \times 10^{11}$ &   0.63 \\  16 & -35.5 & 4.0 &   40 & $  8.4 \times 10^{12}$ & &  $  4.8 \times 10^{11}$ &  17.7 & 0.43 & & $  2.7 \times 10^{11}$  & 31.5 & &  $  1.5 \times 10^{11}$ &   1.38 \\  17 & -41.3 & 3.3 &   40 & $  7.0 \times 10^{12}$ & &  $  7.3 \times 10^{11}$ &   9.6 & 0.24 & & $  4.5 \times 10^{12}$  &  1.6 & &  $  1.6 \times 10^{12}$ &   0.11 \\  18 & -45.7 & 3.0 &   40 & $  4.2 \times 10^{12}$ & &  $  8.0 \times 10^{11}$ &   5.3 & 0.13 & & $  2.6 \times 10^{12}$  &  1.6 & &  $  5.5 \times 10^{11}$ &   0.19 \\  19 & -49.0 & 3.0 &   40 & $  4.2 \times 10^{12}$ & &  $  7.7 \times 10^{11}$ &   5.5 & 0.14 & & $  3.0 \times 10^{12}$  &  1.4 & &  $  2.9 \times 10^{11}$ &   0.36 \\  20 & -53.5 & 4.0 &   20 & $  9.3 \times 10^{12}$ & &  $  9.2 \times 10^{11}$ &  10.0 & 0.49 & & $  1.2 \times 10^{12}$  &  7.9 & &  $  2.0 \times 10^{11}$ &   2.29 \\  21 & -58.7 & 4.0 &   20 & $<  2.8 \times 10^{12}$ & &  $  7.0 \times 10^{11}$ &  $<$  4.0 & $<$ 0.20 & & $  3.2 \times 10^{12}$  & $<$  0.9 & &  $  6.6 \times 10^{11}$ &  $<$  0.21 \\  22 & -69.8 & 6.0 &   20 & $  9.9 \times 10^{12}$ & &  $  2.1 \times 10^{12}$ &   4.8 & 0.23 & & $  1.6 \times 10^{12}$  &  6.1 & &  $  5.1 \times 10^{11}$ &   0.96 \\  23 & -76.0 & 4.5 &   20 & $  9.0 \times 10^{12}$ & &  $  1.6 \times 10^{12}$ &   5.7 & 0.28 & & $  3.3 \times 10^{12}$  &  2.7 & &  $  5.1 \times 10^{11}$ &   0.86 \\  24 & -80.1 & 4.0 &   20 & $  7.0 \times 10^{12}$ & &  $  1.1 \times 10^{12}$ &   6.3 & 0.31 & & $  4.8 \times 10^{11}$  & 14.6 & &  $  2.4 \times 10^{11}$ &   1.43 \\  25 & -84.3 & 5.0 &   20 & $  1.3 \times 10^{13}$ & &  $  2.5 \times 10^{12}$ &   5.0 & 0.24 & & $  9.1 \times 10^{11}$  & 13.9 & &  $  5.1 \times 10^{11}$ &   1.21 \\  26 & -88.5 & 4.0 &   20 & $<  2.8 \times 10^{12}$ & &  $  1.8 \times 10^{12}$ &  $<$  1.6 & $<$ 0.08 & & $  1.5 \times 10^{12}$  & $<$  1.9 & &  $  4.5 \times 10^{11}$ &  $<$  0.31 \\  27 & -93.0 & 4.0 &   20 & $  9.9 \times 10^{12}$ & &  $  1.4 \times 10^{12}$ &   6.9 & 0.34 & & $  2.5 \times 10^{12}$  &  4.0 & &  $  6.0 \times 10^{11}$ &   0.80 \\  28 & -97.5 & 4.0 &   20 & $  1.4 \times 10^{13}$ & &  $  1.1 \times 10^{12}$ &  12.6 & 0.62 & & $  2.0 \times 10^{12}$  &  6.9 & &  $  6.0 \times 10^{11}$ &   1.15 \\  29 & -101.8 & 4.0 &   20 & $  1.1 \times 10^{13}$ & &  $  9.6 \times 10^{11}$ &  11.8 & 0.58 & & $  4.3 \times 10^{12}$  &  2.6 & &  $  1.1 \times 10^{12}$ &   0.48 \\  30 & -106.2 & 4.0 &   20 & $  1.1 \times 10^{13}$ & &  $  4.5 \times 10^{11}$ &  25.2 & 1.24 & & $  4.1 \times 10^{12}$  &  2.8 & &  $  6.0 \times 10^{11}$ &   0.92 \\  31 & -111.6 & 4.0 &   20 & $  8.4 \times 10^{12}$ & &  $<  1.6 \times 10^{11}$ & $>$ 53.0 & $>$ 2.60 & & $  8.6 \times 10^{11}$  &  9.8 & &  $  5.4 \times 10^{11}$ &   0.76 \\  32 & -116.0 & 4.0 &   20 & $  1.4 \times 10^{13}$ & &  $<  1.6 \times 10^{11}$ & $>$ 88.3 & $>$ 4.33 & & $  4.3 \times 10^{11}$  & 32.8 & &  $  4.2 \times 10^{11}$ &   1.64 \\  \hline
 \end{tabular}
 \end{center}
 Notes:
 All components have a rotation temperature of 2.7 K except components $n_c =$ 1 and 2 of c-C$_3$H$_2$ which have 15.0 and  3.0 K, respectively, and component $n_c =$ 2 of H$^{13}$CO$^+$ which has  3.1 K.
 The emission components of the H$^{13}$CO$^+$ model are the same as in Table~\ref{t:hcopmodel} and are not listed here.
 
 $^a$ Component number.
 $^b$ LSR velocity, the same for all four molecules except for 1 component of $^{13}$CH$^+$ (2: 68.7 km s$^{-1}$) 
 , 2 components of c-C$_3$H$_2$ (1: 61.0 km s$^{-1}$, 2: 65.0 km s$^{-1}$) 
 , and 1 component of H$^{13}$CO$^+$ (2: 64.5 km s$^{-1}$). 
 $^c$ Linewidth (\textit{FWHM}), the same for all four molecules except for 2 components of c-C$_3$H$_2$ (1: 15.0 km s$^{-1}$, 2: 11.0 km s$^{-1}$) 
  and for 1 component of H$^{13}$CO$^+$ ( 2: 12.0 km s$^{-1}$). 
  $^d$ Assumed $^{12}$C/$^{13}$C isotopic ratio for HCO$^+$ and CH$^+$.
 $^e$ Column density.
 $^f$ Column density ratio SH$^+$/$^{13}$CH$^+$.
 $^g$ Column density ratio SH$^+$/$^{12}$CH$^+$.
 $^h$ Column density ratio SH$^+$/c-C$_3$H$_2$.
 $^i$ Column density ratio SH$^+$/HCO$^+$.
 \end{table*}

\begin{table*}
 \caption{
 Parameters of the best-fit LTE model of H$^{13}$CO$^+$ and the LTE model of SH$^+$ fitted with the same velocity components as H$^{13}$CO$^+$.
 }
\label{t:hcopmodel}
 \begin{center}
 \begin{tabular}{cccccccc}
 \hline\hline
  \multicolumn{5}{c}{H$^{13}$CO$^+$} & & \multicolumn{2}{c}{SH$^+$}  \\ 
 \cline{1-5} \cline{7-8}
 \noalign{\smallskip}
  \multicolumn{1}{c}{$n_c^a$} & \multicolumn{1}{c}{$V_{\mathrm{lsr}}$$^{b}$} & \multicolumn{1}{c}{$\Delta V^{c}$} & \multicolumn{1}{c}{$N^{d}$} & \multicolumn{1}{c}{$\frac{^{12}\mathrm{C}}{^{13}\mathrm{C}}^e$} & & \multicolumn{1}{c}{$N^{d}$} & \multicolumn{1}{c}{$R_{12}$$^{f}$} \\ 
   & \multicolumn{1}{c}{\scriptsize (km~s$^{-1}$)} & \multicolumn{1}{c}{\scriptsize (km~s$^{-1}$)} & \multicolumn{1}{c}{\scriptsize (cm$^{-2}$)} & & & \multicolumn{1}{c}{\scriptsize (cm$^{-2}$)} & \\ 
 \multicolumn{1}{c}{(1)} & \multicolumn{1}{c}{(2)} & \multicolumn{1}{c}{(3)} & \multicolumn{1}{c}{(4)} & \multicolumn{1}{c}{(5)} & & \multicolumn{1}{c}{(6)} & \multicolumn{1}{c}{(7)} \\ 
 \hline
1 & 62.0 & 12.0 & $ 3.75 \times 10^{14}$ &   20 &  & -- & -- \\ 2 & 80.0 & 20.0 & $ 5.00 \times 10^{13}$ &   20 &  & -- & -- \\ 3 & 104.0 & 20.0 & $ 4.00 \times 10^{12}$ &   20 &  & -- & -- \\ 4 & 51.0 & 12.0 & $ 4.00 \times 10^{13}$ &   20 &  & -- & -- \\ 5 & 64.5 & 12.0 & $ 1.36 \times 10^{14}$ &   20 & & $ 1.97 \times 10^{14}$ & 0.07 \\ 6 & 35.0 & 7.0 & $ 4.82 \times 10^{11}$ &   40 & & $< 2.82 \times 10^{12}$ & $<$ 0.14 \\ 7 & 30.0 & 5.0 & $ 7.83 \times 10^{11}$ &   40 & & $< 2.82 \times 10^{12}$ & $<$ 0.09 \\ 8 & 25.5 & 3.0 & $ 3.01 \times 10^{11}$ &   40 & & $< 2.82 \times 10^{12}$ & $<$ 0.23 \\ 9 & 22.0 & 5.0 & $ 2.53 \times 10^{12}$ &   60 & & $ 8.45 \times 10^{12}$ & 0.05 \\ 10 & 17.0 & 5.0 & $ 2.05 \times 10^{12}$ &   60 & & $ 2.25 \times 10^{13}$ & 0.18 \\ 11 & 12.0 & 6.0 & $ 1.81 \times 10^{12}$ &   60 & & $ 2.82 \times 10^{13}$ & 0.25 \\ 12 & 7.6 & 4.5 & $ 1.20 \times 10^{12}$ &   20 & & $ 2.54 \times 10^{13}$ & 1.03 \\ 13 & 1.1 & 6.2 & $ 4.15 \times 10^{12}$ &   20 & & $ 1.69 \times 10^{13}$ & 0.20 \\ 14 & -5.5 & 7.0 & $ 6.02 \times 10^{11}$ &   20 & & $ 1.97 \times 10^{13}$ & 1.61 \\ 15 & -13.5 & 5.0 & $ 3.01 \times 10^{11}$ &   40 & & $ 1.55 \times 10^{13}$ & 1.26 \\ 16 & -19.5 & 5.0 & $ 6.62 \times 10^{11}$ &   40 & & $ 8.45 \times 10^{12}$ & 0.31 \\ 17 & -26.3 & 5.0 & $ 7.22 \times 10^{11}$ &   40 & & $ 8.45 \times 10^{12}$ & 0.29 \\ 18 & -31.0 & 5.0 & $ 2.17 \times 10^{11}$ &   40 & & $ 8.45 \times 10^{12}$ & 0.96 \\ 19 & -36.5 & 4.0 & $ 1.20 \times 10^{11}$ &   40 & & $ 8.45 \times 10^{12}$ & 1.72 \\ 20 & -41.0 & 3.3 & $ 1.57 \times 10^{12}$ &   40 & & $ 7.04 \times 10^{12}$ & 0.11 \\ 21 & -45.7 & 3.0 & $ 6.02 \times 10^{11}$ &   40 & & $ 4.23 \times 10^{12}$ & 0.17 \\ 22 & -49.5 & 3.0 & $ 3.01 \times 10^{11}$ &   40 & & $ 4.23 \times 10^{12}$ & 0.34 \\ 23 & -54.0 & 4.0 & $ 1.99 \times 10^{11}$ &   20 & & $ 9.30 \times 10^{12}$ & 2.29 \\ 24 & -59.3 & 4.0 & $ 5.72 \times 10^{11}$ &   20 & & $< 2.82 \times 10^{12}$ & $<$ 0.24 \\ 25 & -66.0 & 5.0 & $ 4.21 \times 10^{11}$ &   20 & & $< 2.82 \times 10^{12}$ & $<$ 0.33 \\ 26 & -72.0 & 5.0 & $ 4.21 \times 10^{11}$ &   20 & & $ 9.86 \times 10^{12}$ & 1.15 \\ 27 & -76.3 & 4.0 & $ 4.21 \times 10^{11}$ &   20 & & $ 8.45 \times 10^{12}$ & 0.98 \\ 28 & -80.1 & 4.0 & $ 2.71 \times 10^{11}$ &   20 & & $ 7.04 \times 10^{12}$ & 1.27 \\ 29 & -84.3 & 5.0 & $ 5.12 \times 10^{11}$ &   20 & & $ 1.27 \times 10^{13}$ & 1.21 \\ 30 & -88.5 & 3.5 & $ 4.21 \times 10^{11}$ &   20 & & $< 2.82 \times 10^{12}$ & $<$ 0.33 \\ 31 & -92.6 & 3.5 & $ 5.42 \times 10^{11}$ &   20 & & $ 8.45 \times 10^{12}$ & 0.76 \\ 32 & -97.2 & 4.0 & $ 6.62 \times 10^{11}$ &   20 & & $ 1.41 \times 10^{13}$ & 1.04 \\ 33 & -102.2 & 5.0 & $ 1.57 \times 10^{12}$ &   20 & & $ 1.69 \times 10^{13}$ & 0.53 \\ 34 & -108.0 & 4.0 & $ 3.61 \times 10^{11}$ &   20 & & $ 1.13 \times 10^{13}$ & 1.53 \\ 35 & -112.0 & 4.0 & $ 4.82 \times 10^{11}$ &   20 & & $ 8.45 \times 10^{12}$ & 0.86 \\ 36 & -116.0 & 4.0 & $ 3.61 \times 10^{11}$ &   20 & & $ 1.41 \times 10^{13}$ & 1.91 \\ 37 & -120.0 & 4.0 & $ 1.20 \times 10^{11}$ &   20 & & $ 5.63 \times 10^{12}$ & 2.29 \\ 38 & -123.7 & 4.0 & $ 1.20 \times 10^{11}$ &   20 & & $ 5.63 \times 10^{12}$ & 2.29 \\ 39 & -130.4 & 8.0 & $ 1.99 \times 10^{11}$ &   20 & & $ 7.04 \times 10^{12}$ & 1.74 \\  \hline
 \end{tabular}
 \end{center}
 Notes:
 All components have a rotation temperature of 2.7 K except components $n_c =$ 1 to 4 of H$^{13}$CO$^+$ which have 20.0 K and $n_c =$ 5 of H$^{13}$CO$^+$ which has  3.1 K.
 $^a$ Component number.
 $^b$ LSR velocity, the same for both molecules except for one component of SH$^+$ (5: 64.0 km s$^{-1}$).
 $^c$ Linewidth (\textit{FWHM}), the same for both molecules except for one component of SH$^+$ (5: 19.0 km s$^{-1}$).
 $^d$ Column density.
 $^e$ Assumed isotopic ratio H$^{12}$CO$^+$/H$^{13}$CO$^+$.
 $^f$ Column density ratio SH$^+$/HCO$^+$.
 \end{table*}


The parameters used to fit the SH$^+$ 1$_1$--0$_1$
absorption spectrum are listed in Table~\ref{t:shpmodel} and the fit to the
spectrum obtained with pixel 2 is plotted again in Fig.~\ref{f:3mm}b for a
direct comparison with HCl, $^{13}$CH$^+$, $c$-C$_3$H$_2$, and
H$^{13}$CO$^+$. Although the decision
to fix the linewidths and velocities to those derived from $c$-C$_3$H$_2$
{helps a
lot to constrain the SH$^+$ model, the hyperfine structure of SH$^+$ and the
limited signal-to-noise ratio of the observed spectra do not allow to find a
unique solution. Using the somewhat different set of velocity and linewidth
parameters derived from HCO$^+$, we were able to find a synthetic spectrum for
SH$^+$ as good as the SH$^+$ model derived from $c$-C$_3$H$_2$
(the parameters of this alternative model are listed in
Table~\ref{t:hcopmodel}). Therefore, the
SH$^+$ model described in Table~\ref{t:shpmodel} should be considered as a
possible model rather than the {best-fit model. Nevertheless,
if we assume that $c$-C$_3$H$_2$ and SH$^+$ trace the same diffuse clouds, then
it makes sense to compare the column densities listed in
Table~\ref{t:shpmodel}. Column 10 lists the column density ratio
SH$^+$/$c$-C$_3$H$_2$. It varies by nearly two orders of magnitude, from $<0.6$
to $\sim 33$. As a caveat, we remind that the SH$^+$ model assumes no
contamination from emission lines of other molecules. If there are
contaminating emission lines, then the column densities derived for the SH$^+$
components contaminated by these emission lines are lower limits. 

\begin{figure}[t]
\centerline{\resizebox{0.8\hsize}{!}{\includegraphics[angle=270]{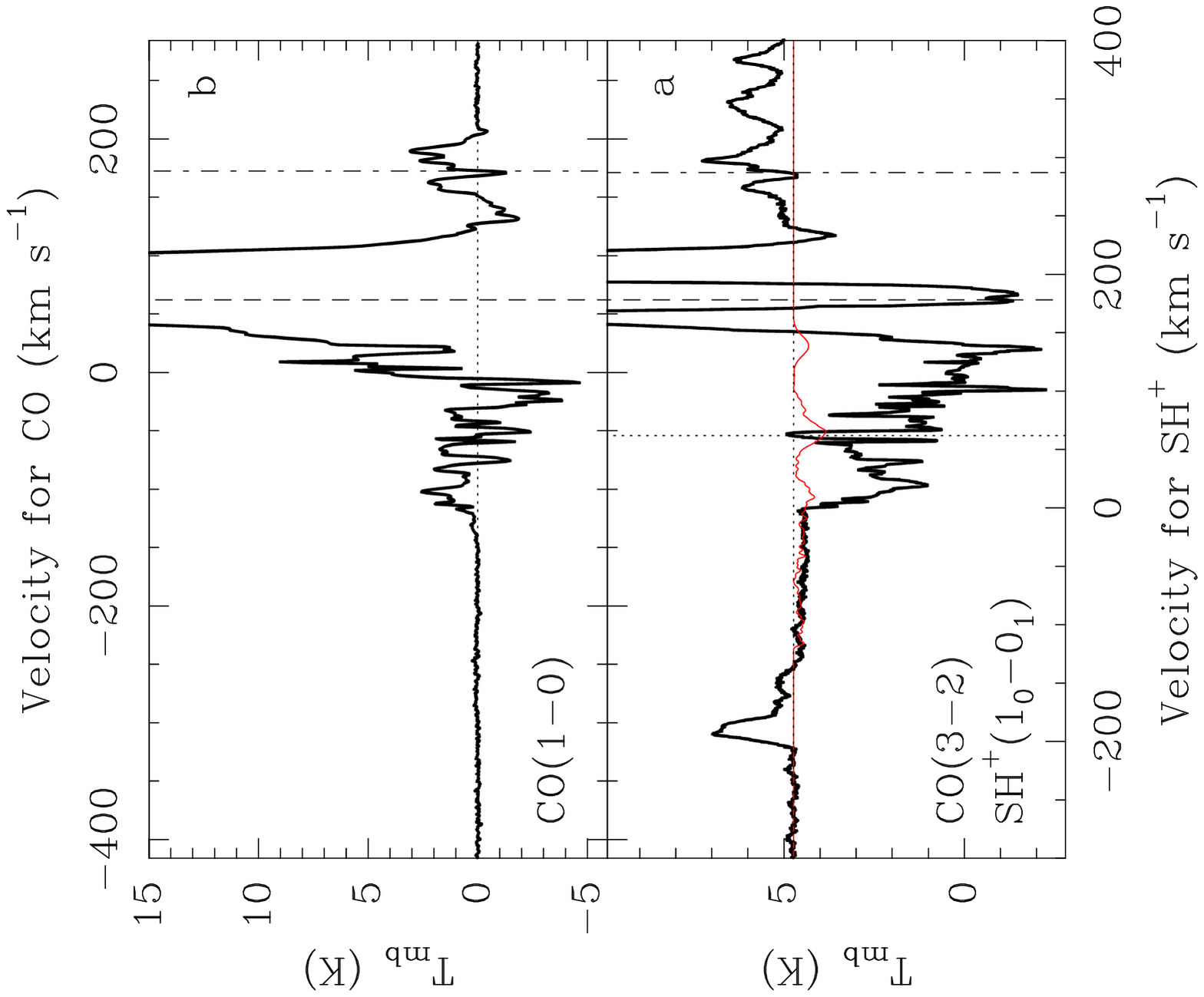}}}
\caption{
\textit{a} Spectrum obtained toward Sgr~B2(M) at the frequency of
the SH$^+$ 1$_0$-0$_1$ line with the APEX telescope in main beam brightness
temperature. The red curve is the synthetic spectrum described in
Sect.~\ref{ss:modb2m}, the horizontal dotted line shows the continuum level,
and the vertical dotted line marks the systemic velocity of the source
(62~km~s$^{-1}$). The strong absorption lines are spiral arm components of
CO 3--2. The lower axis gives the LSR velocity for SH$^+$, while the
upper axis refers to CO, with the same labeling as in panel \textit{b}.
\textit{b} CO(1-0) spectrum obtained toward Sgr~B2 as part of our
line survey with the IRAM 30~m telescope. The horizontal dotted line indicates the zero
level (after baseline removal because the observations were performed in
position-switching mode with a far OFF position, which yields a very uncertain
baseline level). In both panels, the dashed line
marks the systemic velocity of the source in CO 3--2 and the dot-dashed line
indicates the CO velocity +172.6 km~s$^{-1}$ at which an extra absorption component 
appears in the $^{13}$CH$^+$ spectrum (see discussion in \S\ref{13chplus}.
}
\label{f:shp345}
\end{figure}

\begin{figure}[t]
\centerline{\resizebox{0.8\hsize}{!}{\includegraphics[angle=270]{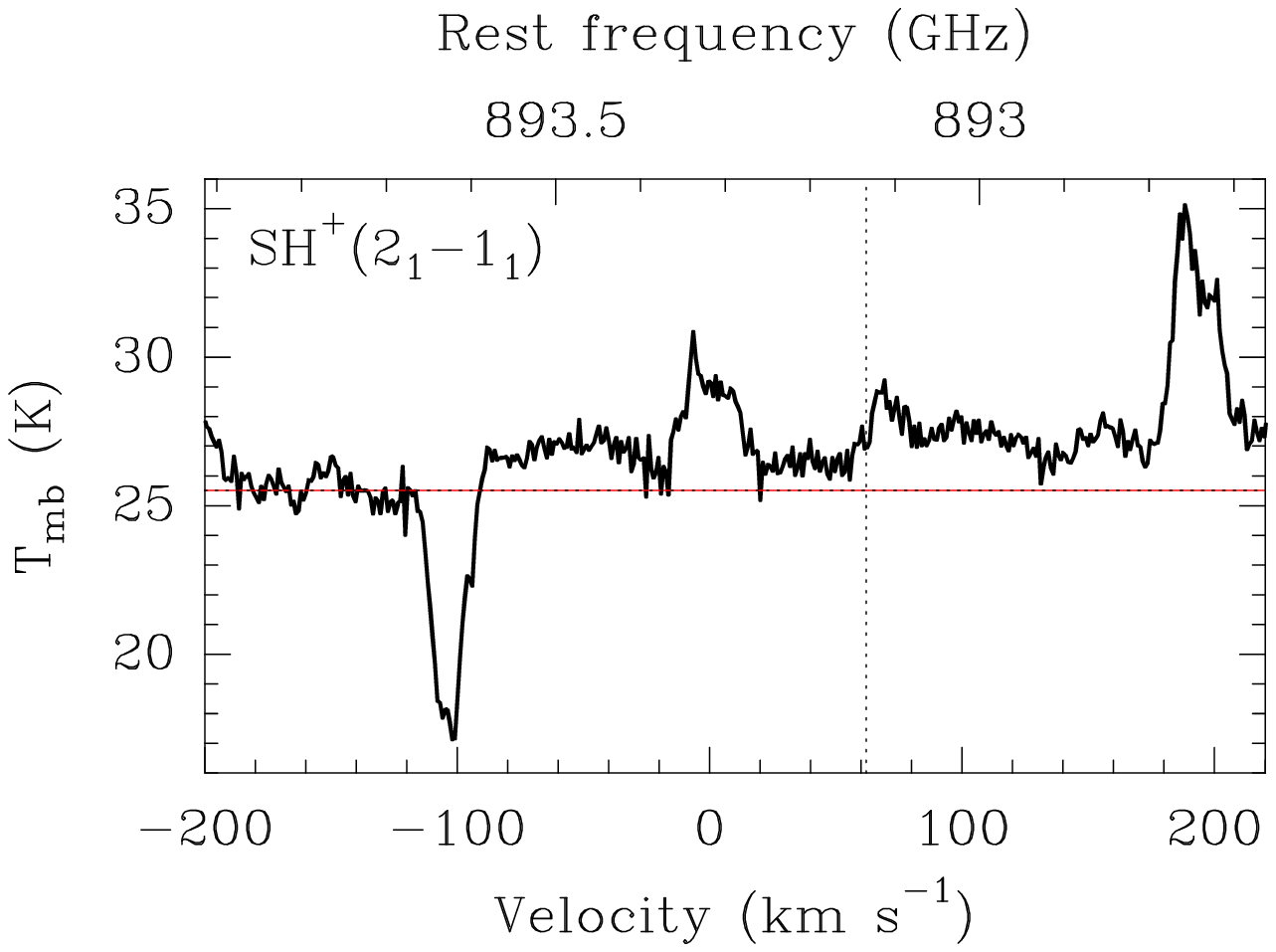}}}
\caption{
Spectrum obtained toward Sgr~B2(M) at the frequency of
SH$^+$ 2$_1$-1$_1$ with the APEX telescope in main beam brightness
temperature. The red curve is the synthetic spectrum described in
Sect.~\ref{ss:modb2m}, i.e., showing no detectable emission, the horizontal dotted line shows the continuum level,
and the vertical dotted line marks the systemic velocity of the source
(62~km~s$^{-1}$). The absorption line appearing at $\sim -105$ km~s$^{-1}$ 
is the  $J_{K_{a}K_{c}} = 1_{11} - 0_{00}$ ground-state line of HDO at
893638.7 MHz \citep{Comito2003}.
}
\label{f:shp893}
\end{figure}

Unfortunately, the SH$^+$ 1$_0$--0$_1$ line at 345.9 GHz is heavily blended
with the spiral arm components of the CO 3--2 line seen in absorption (see
Fig.~\ref{f:shp345}a). The $T_{\rm rot}$ $= T_{\rm CMB}$  model derived 
for the 1$_1$--0$_1$ line predicts 1$_0$--0$_1$ spiral arm components 
in absorption that are relatively consistent with the observed spectrum 
dominated by the CO 3--2 features, except around 62~km~s$^{-1}$, 
the systemic velocity of Sgr~B2(M). At 893 GHz, there is no clear
evidence for the SH$^+$ 2$_1$--1$_1$ line (see Fig.~\ref{f:shp893}). 
Our $T_{\rm rot}$ $= T_{\rm CMB}$ model predicts no significant absorption 
for this higher energy transition (see red spectrum in Fig.~\ref{f:shp893}). 
On the other hand, this transition could partly contribute 
to the broad feature detected in emission between 58 and 125~km~s$^{-1}$ 
(although the higher velocity half of that feature is uncertain due
to the limited baseline quality). An additional model component in emission is
indeed permitted by the 893 GHz spectrum and can even solve the disagreement at
62 km~s$^{-1}$ between the synthetic and observed spectra of the 1$_0$--0$_1$
transition. If this emission component has a size smaller than $\sim
20\arcsec$, then the $T_{\rm rot}$ $= T_{\rm CMB}$  synthetic spectrum of the 
($-20.4'',+1.4")$ CHAMP+ pixel at 683 GHz (shown in Fig.~\ref{f:3mm}) will be 
unaffected and still provide a good fit to the observations. 
However, the fit to the spectrum of the central pixel at 683 GHz (see Fig.~\ref{f:shp}) 
will be worse, and a more elaborate radiative transfer modeling would be needed 
to find a consistent fit for this potential emission component.

\subsubsection{\label{13chplus}The  $^{13}$CH$^+$ absorption}
The $^{13}$CH$^+$ 1--0 transition detected in absorption toward Sgr~B2(M) was
also modeled in the $T_{\rm rot}$ $= T_{\rm CMB}$  approximation (see Fig.~\ref{f:3mm}c). 
We used the SH$^+$ model as a reference, keeping the velocity offsets and 
linewidths unchanged. The resulting fit, shown in red in this figure, 
is in good agreement with the observed spectrum, although it could certainly 
be somewhat better if the line widths were taken as free parameters for the
fitting. 

The column densities resulting from this fit are listed in
Table~\ref{t:shpmodel}, as well as the SH$^+$/$^{13}$CH$^+$ 
and the SH$^+$/$^{12}$CH$^+$ column density
ratio.  The latter varies significantly from component to component,
from $0.07$ to 1.6, 
if we exclude component 32 in  Table~\ref{t:shpmodel} at whose velocity the
signal-to-noise ratio in the SH$^+$ spectrum is very low and which is not detected in the 
$^{13}$CH$^+$ absorption at all. 

The $^{13}$CH$^+$ $1-0$ spectrum shows an absorption feature at
$+172.6$~ km~s$^{-1}$.
%
Although absorption in CO is detected at this velocity in our
 3~mm line survey in the CO 1--0 and $^{13}$CO 1--0 lines (see
 Fig.~\ref{f:shp345}), this component is neither detected in
 H$^{13}$CO$^+$ 1--0, $c$-C$_3$H$_2$ 2$_{12}$--1$_{01}$, nor in
 SH$^+$ 1$_1$--0$_1$ (see Fig.~\ref{f:3mm}). Features at this velocity are
 usually attributed to the Expanding Molecular Ring
 \citep[EMR, e.g.][]{Morris1996} but $+172.6$~ km~s$^{-1}$ would refer
 to the far side of the ring, hence no absorption is expected and it
 is possibly more likely that this feature is an as yet unidentified
 (U) absorption line.
%

We also list in Table~\ref{t:shpmodel} the parameters of an alternative 
H$^{13}$CO$^+$ $T_{\rm rot}$ $= T_{\rm CMB}$  model fitted with the same velocity
components as $c$-C$_3$H$_2$ (i.e. same linewidths and LSR velocities). This
model is less good than the one that we described above and that we showed in
Fig.~\ref{f:3mm}e, but it can be used to roughly estimate the SH$^+$/HCO$^+$
column density ratio (Col.~12 of Table~\ref{t:shpmodel}). 
This ratio varies by
about one order of magnitude, from 0.07 to 2.38.

\subsubsection{\label{sss:hcl}Modeling the HCl absorption only from the Sgr~B2 region}
The HCl and H$^{37}$Cl transitions detected in absorption 
toward Sgr~B2(M) were also modeled in the $T_{\rm rot}$ $= T_{\rm CMB}$ approximation 
(see Fig.~\ref{f:hcl}). We used the SH$^+$ model as a reference, 
keeping the velocity offsets and linewidths unchanged, except for the main absorption 
component associated with Sgr~B2(M) itself. 
The resulting fit is shown in red in Fig.~\ref{f:hcl} and the fit parameters 
are listed in Table~\ref{t:hclmodel}. Only the main absorption component is 
clearly detected for both isotopologues. 
For all other velocity components but one, we used a column density of 
$2.8 \times 10^{12}$ cm$^{-2}$, represented by the red line in Fig.~\ref{f:hcl},  to get a rough estimate of the
column density upper limits. Only one additional velocity component ($n_c = 6; \sim17$ km~s$^{-1}$)
marginally shows absorption in the main isotopologue HCl at 2.8 times this value.
Due to the compact hyperfine structure, the line width of the main absorption 
component (8 km~s$^{-1}$) is somewhat uncertain, but our modeling suggests 
that it is significantly smaller than for the other species. 
Finally, for Sgr B2(M) we derive an isotopic [H$^{35}$Cl$/$H$^{37}$Cl] ratio of 
$\sim 4$, consistent with the terrestrial value.

\begin{figure}[t]
\centerline{\resizebox{\hsize}{!}{\includegraphics[angle=270]{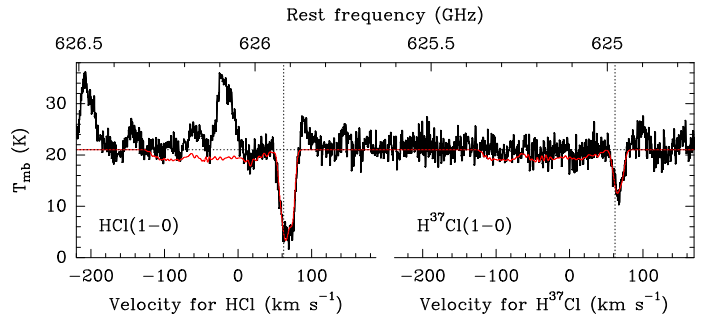}}}
\caption{
Spectrum obtained toward Sgr~B2(M) around 625 GHz with the APEX telescope in 
main beam brightness temperature. The red curve is the synthetic spectrum 
described in Sect.~\ref{ss:modb2m}, the horizontal dotted line shows the 
continuum level, and the left and right vertical dotted lines mark the 
systemic velocity of the source (62~km~s$^{-1}$) for HCl and H$^{37}$Cl, 
respectively. The strong emission close the HCl is likely due to a SO$_2$
line at 626087.3~MHz.
}
\label{f:hcl}
\end{figure}

\begin{table}
 \caption{
 Parameters of the best-fit LTE model of HCl and H$^{37}$Cl.
 }
\label{t:hclmodel}
 \begin{center}
 \begin{tabular}{ccccc}
 \hline\hline
 \noalign{\smallskip}
 \multicolumn{1}{c}{Molecule} & \multicolumn{1}{c}{$n_c^a$} & \multicolumn{1}{c}{$V_{\mathrm{lsr}}$$^{b}$} & \multicolumn{1}{c}{$\Delta V^{c}$} & \multicolumn{1}{c}{$N^{d}$} \\ 
   & & \multicolumn{1}{c}{\scriptsize (km~s$^{-1}$)} & \multicolumn{1}{c}{\scriptsize (km~s$^{-1}$)} & \multicolumn{1}{c}{\scriptsize (cm$^{-2}$)} \\ 
 \multicolumn{1}{c}{(1)} & \multicolumn{1}{c}{(2)} & \multicolumn{1}{c}{(3)} & \multicolumn{1}{c}{(4)} & \multicolumn{1}{c}{(4)} \\ 
 \hline
HCl & 2 & 65.0 & 8.0 & $ 1.90 \times 10^{14}$ \\  & 6 & 17.0 & 5.0 & $ 7.60 \times 10^{12}$ \\ \smallskip 
H$^{37}$Cl & 2 & 65.0 & 8.0 & $ 4.75 \times 10^{13}$ \\  \hline
 \end{tabular}
 \end{center}
 Notes:
 $^a$ Same component numbering as in Table~\ref{t:shpmodel}.
 $^b$ LSR velocity.
 $^c$ Linewidth.
 $^d$ Column density.
 \end{table}

\subsection{Summary of modeling}
To visualize the results of our modeling presented in Tables~\ref{t:shpmodel}
and \ref{t:hcopmodel}, we show, in Fig.~\ref{f:chpshp}, the column densities 
and abundances that we derived for the species considered functions of LSR velocity.  

\section{Discussion}
\label{discussion}
\subsection{Column densities and abundances}
\subsubsection{Caveats}
We would like to precede our discussion of the SH$^+$,  $^{13}$CH$^+$, and HCl
abundances in the spiral arm clouds with some cautionary notes. 
First, the assignment of velocity features to Galactocentric distance 
is by no means clear-cut. 
For example, \citet{Sofue2006}, from a careful analysis of the 
longitude--velocity diagram of the GC region, assigns certain velocity features 
to peculiar components of the complex Galactic center velocity field, 
namely the GC molecular ring and the GC expanding ring, both within a 
Galactocentric radius, $R$, of 600 pc of the GC. In addition to several spiral arms, 
he invokes expanding rings at $R = 3$ and 4 kpc. Some of Sofue's conclusions are 
at variance from the LSR velocity--location assignment we present in 
\S\ref{ss:modb2m}.  Second, SH$^+$ and $^{13}$CH$^+$ may in principle be
associated with predominantly atomic or molecular diffuse clouds. 
For meaningful abundance calculations, it is thus necessary to have values 
for 
both
the atomic hydrogen column density, $N({\rm HI})$, 
and the molecular hydrogen column density, $N({\rm H}_2)$. 
Actually, the clouds from which we see absorption may have complex structure, 
e.g., a Giant Molecular Cloud  region and a diffuse atomic envelope (see \S\ref{origin}) and/or 
we may be dealing with more than one cloud along each line of sight. 
Certainly, the HI spin 
temperatures
are in the range of values 
found for diffuse clouds and significantly higher than the canonical GMC values of 20--30 K. 

Two final, more technical caveats relate to the way we derive column densities. 
For the 3~mm Sgr~B2(M) spectra for H$^{13}$CO$^+$ and c-C$_3$H$_2$ 
we have constructed full XCLASS models considering all the absorption 
from these molecules \textit{plus} emission from them and all other molecules. 
In contrast, for the submillimeter spectra, we have not done this. 
While even the 3 mm spectra may be contaminated by emission features 
inadequately treated by the model, for the submm absorption spectra 
this is a much greater source of uncertainty for line assignments and 
intensity predictions, making, strictly speaking all the column densities 
we determine lower limits. 
We do at present not feel confident to construct a comprehensive model 
for the submm spectra. 
Therefore, we have just marked the positions of stronger lines 
in the spectra by the species from which the originate. 
Similarly, if, as discussed in \S \ref{analysis}, excitation by the ambient radiation field 
is non-negligible, then  $T_{\rm rot} > T_{\rm CMB}$, 
which would make our column densities underestimates.

Finally, our XCLASS-based method makes it difficult to derive statistical errors 
and, thus, upper limits rigorously in a $\chi^2$ sense. 
Rather, upper limits are determined by assuming a column density, calculating a model spectrum 
and then ''eye-balling'' whether it is consistent with the data.

\subsubsection{Atomic and molecular hydrogen column densities}
\label{hicolumn}
The column density of atomic hydrogen, $N({\rm HI})$, can be calculated 
from measurements of the 21~cm HI line's optical depth, $\tau({\rm HI})$, 
at the intervening clouds' velocities \citep[e.g., ][]{GarwoodDickey1989} 
and a determination of their spin temperatures, $T_{\rm }s$, as described 
by \citet{Lazareff1975}. 
The method invokes a combination of interferometric observations to, first, 
directly determine the optical depths of the absorption components 
and, second, single dish observations to measure the (optically thick) 
line's brightness temperature. 
As described in a forthcoming paper, we have performed the necessary 
single dish observations with  the Effelsberg 100 meter telescope 
and obtained an optical depth spectrum ($\tau({\rm HI})$
vs. $V_{lsr}$) measured toward Sgr~B2(M) with the NRAO Very Large Array 
from C. Lang (pers. comm.).  
We note our values for $N({\rm HI}$ are generally lower 
than those derived by \citet{Vastel2002}. 
This is partially due to their assumption of $T_{\rm s} =150$~K for all velocity components.

A determination of $N({\rm H}_2)$ is more difficult, as selection 
of a ''universal'' trace molecule for molecular hydrogen is not straightforward. 
For meaningful modeling, such a molecule's lines would be required 
to be subthermally excited, meaning that their relative level populations 
be determined by $T_{\rm CMB}$. 
This means, for example that, due to their small dipole moments 
and resulting low critical densities, isotopologues of CO are not suited. 
A possible choice for an H$_2$ tracer, which we adopt here, 
might be HCO$^+$. 
\citet{LucasLiszt1996} from their study of this molecule's 89~GHz 
$J =1-0$ ground-state transition, conclude its abundance 
(relative to H$_2$) to be ''$\sim3$--$6 \times 10^{-9}$ across 
a very broad range of extinction'', meaning between hydrogen column densities 
$\sim 0.1$ and $4 \times 10^{20}$~cm$^{-2}$. 
These authors base this finding, first, on the very tight correlation 
between the HCO$^+$ and the OH column density and, 
second, on the fact that OH has the same abundance ($10^{-7}$) 
in diffuse as in dense dark clouds. 
While their own absorption data only sample hydrogen column densities 
between $10^{20}$ and $2 \times 10^{21}$~cm,$^{-1}$, 
they find that the correlation holds at 200 times higher values.
Here we adopt $5 \times 10^{-9}$ for the HCO$^+$/H$_2$ abundance ratio. 
This value is of the same magnitude as that listed by \citet{vanDishoeck1993prpl} 
for (high column density) regions of low and high mass star formation.

Since HCO$^+$ absorption toward Sgr~B2 is optically thick even for the 
non-Sgr~B2 region velocity features, we use, as described in \S\ref{ss:modb2m}, 
H$^{13}$CO$^+$ spectra and obtain the main-isotope's column density 
by scaling with the appropriate $^{12}$C/$^{13}$C ratios 
given in column~4 of Table~\ref{t:shpmodel}. 

\begin{table*}
 \caption{
 \ion{H}{i} and H$_2$ column densities of the absorption components.
 }
\label{t:hi}
 \begin{center}
 \begin{tabular}{ccccccc}
 \hline\hline
 \noalign{\smallskip}
 \multicolumn{1}{c}{Label} & \multicolumn{1}{c}{Velocity range$^a$} & \multicolumn{1}{c}{$N_{\mathrm{\ion{H}{i}}}/T_s$$^b$} & \multicolumn{1}{c}{$T_{\rm s}$$^c$} & \multicolumn{1}{c}{$N_{\mathrm{\ion{H}{i}}}$$^d$} & \multicolumn{1}{c}{$N_{{\mathrm{H}}_2}$$^e$} & \multicolumn{1}{c}{Abs. comp.$^{f}$} \\ 
   &\multicolumn{1}{c}{\scriptsize (km~s$^{-1}$)} & \multicolumn{1}{c}{\scriptsize (cm$^{-2}$ K$^{-1}$)} & \multicolumn{1}{c}{\scriptsize (K)} & \multicolumn{1}{c}{\scriptsize (cm$^{-2}$)} & \multicolumn{1}{c}{\scriptsize (cm$^{-2}$)} & \\ 
 \multicolumn{1}{c}{(1)} & \multicolumn{1}{c}{(2)} & \multicolumn{1}{c}{(3)} & \multicolumn{1}{c}{(4)} & \multicolumn{1}{c}{(5)} & \multicolumn{1}{c}{(6)} & \multicolumn{1}{c}{(7)} \\ 
 \hline \\[-2.0ex]
A &  95.3 --  38.6 & $  9.0( 1.8) \times 10^{19}$ &  43( 5) & $  3.9( 0.9) \times 10^{21}$ & $  5.4 \times 10^{23}$ & [2] \\ 
B &  23.2 --  12.9 & $  1.5( 0.5) \times 10^{19}$ &  65(10) & $  1.0( 0.4) \times 10^{21}$ & $  5.5 \times 10^{22}$ & [5, 6] \\ 
C & -12.9 -- -33.5 & $  8.5( 2.0) \times 10^{18}$ &  93(10) & $  7.9( 2.0) \times 10^{20}$ & $  1.4 \times 10^{22}$ & [12, 13, 14, 15] \\ 
D & -36.1 -- -51.5 & $  1.6( 0.4) \times 10^{19}$ &  20( 3) & $  3.1( 0.9) \times 10^{20}$ & $  2.0 \times 10^{22}$ & [17, 18, 19] \\ 
E & -54.1 -- -64.4 & $  3.0( 1.0) \times 10^{18}$ &  55( 6) & $  1.7( 0.6) \times 10^{20}$ & $  2.6 \times 10^{21}$ & [21] \\ 
F & -69.6 -- -85.0 & $  2.6( 0.7) \times 10^{18}$ &  80( 8) & $  2.1( 0.6) \times 10^{20}$ & $  7.1 \times 10^{21}$ & [22, 23, 24, 25] \\ 
G & -90.2 -- -123.7 & $  2.8( 0.7) \times 10^{18}$ &  43( 7) & $  1.2( 0.4) \times 10^{20}$ & $  1.6 \times 10^{22}$ & [27, 28, 29, 30, 31, 32] \\ 
 \hline
 \end{tabular}
 \end{center}
 Notes: the numbers in parentheses are the uncertainties.
 $^a$ Velocity range over which the \ion{H}{i} column density is computed.
 $^b$ Ratio of \ion{H}{i} column density to spin temperature, obtained from the VLA data.
 $^c$ \ion{H}{i} spin temperature derived from the Effelsberg data.
 $^d$ \ion{H}{i} column density.
 $^e$ H$_2$ column density derived from the H$^{13}$CO$^+$ column densities assuming the isotopic ratios listed in Col. 4 of Table~\ref{t:shpmodel} and an HCO$^+$ abundance of $ 5 \times 10^{-9} $ relative to H$_2$.
 $^g$ List of the absorption components used to compute the H$_2$ column densities, with the same numbering as in Col. 1 of Table~\ref{t:shpmodel}.
 \end{table*}

Our estimates for  $N({\rm HI})$ and  $N({\rm H}_2)$ for the absorption components 
are listed in Table \ref{t:hi}. For simplicity's sake and motivated by our 
optical depth fitting of the HI data and also to allow 
comparison with other studies, we have grouped the velocity components into seven intervals. 
The HI and H$_2$ column densities and HI spin 
temperatures were derived as described above from 
HI and H$^{13}$CO$^+$  data 
fixed to values compatible with the values given in the table. 
The column densities of SH$^+$ and CH$^+$, 
presented in Table \ref{t:ratios}, 
were obtained by summing over the individual velocity components 
in Table~\ref{t:shpmodel} and multiplying the CH$^+$ column densities by the $^{12}$C/$^{13}$C ratios 
given for each component in that table. 

\subsection{\label{origin}The origin of the $^{13}$CH$^+$ and SH${^+}$ absorption}

Since it is a priori not clear whether the observed SH$^+$, CH$^+$ and HCl molecules 
are only associated with either atomic or molecular gas or both, when calculating abundance ratios, 
we might want to refer to the total column density of hydrogen nuclei, 
$N({\rm H}) \equiv N({\rm HI}) + 2~N({\rm H}_2)$.  
Inspecting Table \ref{t:hi}, we find that over all velocity ranges showing 
absorption, $N({\rm H}_2$) greatly exceeds  $N({\rm HI}$), by factors between 13 and 133, 
approaching the value for Sgr~B2 (138). 
This means that the absorbing medium is dominated by GMC gas. 
This is reinforced by the densities derived by \citet{Greaves1992} 
from LVG modeling of the two lowest rotational lines of CS, which, between 0.4 
and $2 \times 10^4$~cm$^{-2}$, are actually even higher than values in most of a GMC's volume. 

Both the total hydrogen nuclei column density, $N({\rm H})$, and the H$_2$/HI 
ratio are much larger than the values  found in translucent and diffuse clouds.
\citet{Gredel1992} in their study of the archetypical translucent cloud toward HD 210121 
find $N({\rm H}_2) = 8 \times 10^{20}$~cm$^{-2}$ and 
$N({\rm HI}) = 1 - 5 \times 10^{20}$~cm$^{-2}$ for the molecular and 
atomic hydrogen absorption column densities, respectively. 
In comparison, the classic diffuse cloud line of sight toward $\zeta$~Oph 
has comparable HI and H$_2$ column densities of $\approx 5 \times 10^{20}$~cm$^{-2}$ 
or a column density of hydrogen nuclei of 
$N({\rm H}) = 1.5 \times 10^{21}$~cm$^{-2}$ \citep{vanDishoeckBlack1986}.


As discussed above, the absorbing material is dominated by GMCs. 
There is, however, nothing that does imply that the observed CH$^+$ 
and SH$^+$ is associated with gas within a  GMC, although little is known 
about the abundance of both species in such objects. 
Model calculations predict significant column densities in lower-$A_V$ 
portions of dense  
($>10^5$ cm~s$^{-3}$) PDRs illuminated by intense ultraviolet fields \citep{SternbergDalgarno1995}. 
Even more extreme conditions (XDRs) 
are required to produce significant amounts of SH$^+$ in dense regions \citep{Abel2008}. 
However, both dense P and XDRs have small volume filling factors, 
making them unlikely absorbers.
These findings point to an (at least) two component picture 
for most of the absorption components, 
i.e., a diffuse envelope traced by HI,  CH$^+$ and SH$^+$ absorption  plus 
a denser core region, a view that is also invoked by \citet{Vastel2002} 
to explain the absorption they observed with ISO in the C$^+$ 
and O$^0$ fine structure transitions.

In Fig. \ref{f:chp-plot}} we plot  CH$^+$ column density vs. hydrogen column density, 
$N({\rm HI})$ as defined above, 
for the seven velocity ranges in Table \ref{t:hi} (coded A--G). 
For each  CH$^+$ column density 
entry two data points are plotted, both taken from the table. 
The left one (open squares) assumes that the 
CH$^+$ is only associated with atomic hydrogen,  where the right one 
(filled squares) assumes it is only associated with molecular hydrogen or, 
essentially, 
\textit{all} the hydrogen since $N({\rm H_2}) \gg N({\rm HI})$ 
for all velocities. We also plot (as circles) data 
for the two classic diffuse clouds against $\zeta$ Oph and $\zeta$ Per. 
For these, $N({\rm H_2}) \approx  N({\rm HI}) \approx 5 \times 10^{20}$~cm$^{-2}$. 
In addition, we present  data for the translucent cloud toward HD~210121 where, 
again, two points assuming 
that the CH$^+$ is only associated with atomic (left) or molecular gas (right) are plotted. 
All values for the latter three sources, $N({\rm H_2})$, $N({\rm HI})$, and 
$N({\rm CH}^+)$,  are taken from \citet{deVries1988}. 
The circles represent the absorption data 
\citet{Gredel1997} determined toward  a sample of southern OB associations. 
For these lines of sight, which have higher extinctions, 
between 0.36 and 4.2, (mostly) corresponding to translucent clouds, 
we cannot discriminate between atomic and molecular hydrogen columns 
as Gredel lists $N({\rm CH})^+$ vs. $A_V$, which contains contributions from both. 
To convert his $A_V$ values to hydrogen column densities 
we use the results of the  classic study of \citet{Bohlin1978}, 
who found for low values of visual extinction ($0.02 < A_V < 0.6$) a linear relation 
between $A_V$, and hydrogen column density, which as first argued 
by \citet{Dickman1978}, also holds for significantly higher $A_V$ values. 
It is consistent with a relation derived 
from modeling of x-ray data \citep{Guver2009}: 
$N({\rm H}({\rm cm}^2)) \approx 2 \times 10^{21}~A_V({\rm mag}).$

\subsubsection{Comparison with individual diffuse clouds}
While we advocate above a diffuse cloud origin of the CH$^+$ absorption, 
we note that, apart from one velocity range (B, 12.9--23.2~km~s$^{-1}$), 
our CH$^+$ column densities are much higher than found in diffuse clouds 
and mostly higher than in the translucent cloud sample. 
However, the velocity ranges over which the values were determined, 
are mostly much larger than the line widths found in diffuse 
or translucent clouds (and even GMCs). 
For this reason, we plot in the lower panel of Fig.~\ref{f:chp-plot} 
the average column densities that a cloud with ``typical'' line width 
$\Delta V_{\rm  typ}$ would have. 
This means we have divided for each of the velocity intervals A--G, both, 
CH$^+$ and hydrogen column densities by the ratio of the width 
of the velocity interval and $\Delta V_{\rm  typ}$. 
We chose $\Delta V_{\rm  typ}$ to be 4.3 km~s$^{-1}$ FWHM, which is the mean 
of the sample of diffuse lines of sight for which \citet{Crane1995} 
present high spectral resolution CH$^+$ data.\footnote{The Doppler parameter, $b$, 
listed by this and numerous other optical absorption studies and FWHM 
line width are related via $\Delta V_{1/2} = 2\sqrt{ln 2}~b \approx 1.665~b$.} 
This value is consistent with that discussed by \citet{Gredel1997} 
for his translucent cloud sample. 
The thus  calculated values correspond to a ``typical'' cloud with averaged properties.

Inspecting the CH$^+$  column densities for these typical clouds 
we find values very similar to those found in diffuse and translucent clouds. 
We note that the ratio of molecular to atomic hydrogen column densities is 
in most cases higher than that found in translucent and much higher 
than in diffuse molecular clouds. 
The reason for this is almost certainly sight line superposition 
of the CH$^+$-containing diffuse/translucent and denser molecular clouds, 
as discussed above.


\begin{figure}[t]
\centerline{\resizebox{0.8\hsize}{!}{\includegraphics[angle=0]{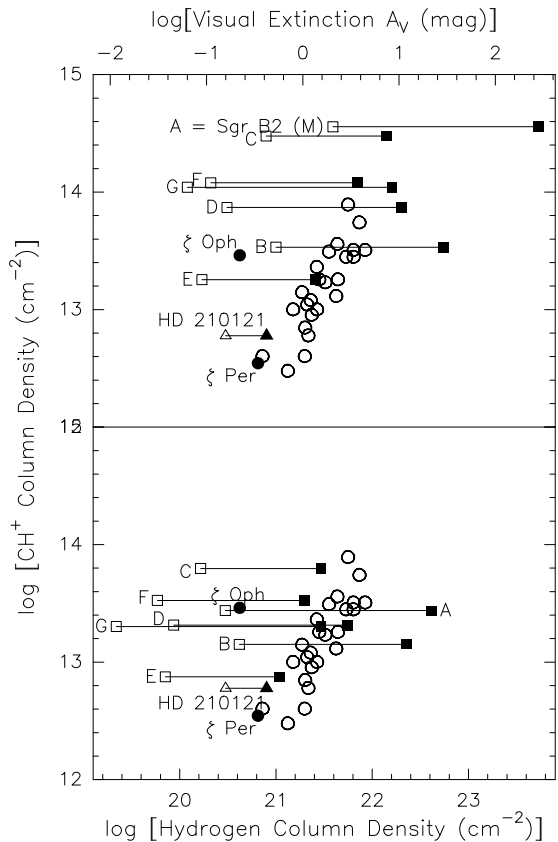}}}
\caption{CH$^+$ column density versus hydrogen column density. 
The ordinate gives the CH$^+$  column density for the seven different velocity intervals 
listed in Table \ref{t:ratios}. The capital letters designate the entries 
in table (top to bottom). The open circles  represent data 
taken from the diffuse/translucent cloud sample of \citet{Gredel1997}. 
The squares connected by lines mark the CH$^+$ column densities 
determined in this work. 
For the open  squares it is assumed that all the absorption 
is from atomic gas; the abscissa corresponds to the HI column density 
(Table~\ref{t:hi}), whereas for the filled squares it is assumed 
that all the absorption is from molecular gas; the abscissa corresponds 
to the H$_2$ column density (Table \ref{t:hi}). 
For the upper panel, column densities were calculated 
from the absorption over the whole velocity range of each interval, 
whereas in the lower panel average column densities normalized 
to a typical line width of 4.3 km~s$^{-1}$ were used (see text).}
\label{f:chp-plot}
\end{figure}

\subsection{Comparison -- SH$^+$ and other S-bearing species and the CH$^+$/SH$^+$ ratio}
%
The values we derive for the SH$^+$ column density, $N({\rm SH}^+)$,
for the non-Sgr~B2 region velocity ranges go up to $7 \times
10^{13}$~cm$^{-2}$ (see Table \ref{t:ratios}).  Considering again the
differences in line widths, these values are comparable with the upper
limits found for optical SH$^+$ absorption from diffuse and
translucent clouds ( i.e. $1 \times 10^{13}$~cm$^{-2}$; see
\S\ref{shpintro}).

The column densities of other sulfur bearing species in the velocity
ranges of Table~\ref{t:ratios} have been determined by
\citet{Greaves1992}, who studied CS, and \citet{Tieftrunk1994} (SO,
H$_2$S, and SO$_2$).  Comparing with our values for SH$^+$ (Table
\ref{t:hi}), we find an [SH$^+$/CS] abundance ratios between 0.5 and
3, [SH$^+$/SO] between 1.6 and $>55$, and [SH$^+$/SO$_2 ] > 2$.  For the
[SH$^+$/H$_2$S] ratio we find values $>2$, except for the $-33.2$ to
$-50.1$ km~s$^{-1}$ velocity range, for which we find the low ratio
of 0.3.  We note that the velocity of this absorption places
it in the Norma spiral arm.  From Table \ref{t:hi} we see that this
range has the highest H$_2$ column density and the highest
H$_2$/HI ratio of all the (non-Sgr~B2 region) absorption components
and resembles most closely a GMC or GMC core.  This reinforces
SH$^+$'s nature as a tracer of more diffuse gas whose abundance drops
in denser environments.

As discussed by \citet{Tieftrunk1994}, SH$^+$ and H$_2$S are chemically related, 
as H$_3$S$^+$ , from which the latter forms via dissociative recombination 
is formed by the reaction
$$
{\rm SH}^+ +{\rm H}_2 \rightarrow {\rm H}_3{\rm S}^+ + {\rm h}\nu  .
$$

With respect to chemical models \citep{Millar1986, Pineau1986}, 
it is also worth noting that, as can be seen from Table~\ref{t:ratios}, 
the SH$^+$/CH$^+$ ratio, [SH$^+$/CH$^+$], which is 0.56  for Sgr~B2(M) 
only varies between 0.11 and 0.26 for all but one of the 
non-Sgr~B2 absorption features. These values are consistent with 
the upper limits on this ratio (0.12--2) 
for diffuse and translucent lines of sight (see \S\ref{shpintro}). 
The component with the higher ratio, $\approx 0.9$ is in the $-90.2$ to 
$-123.7$~km~s$^{-1}$ range, which correspond to peculiar velocity gas 
associated with the GC. 
Interestingly, this velocity range has also by far the highest 
SiO column density measured for the various velocity components by \citet{Greaves1996}. 
These two findings may have a common (shock-wave) origin as an increased 
SiO abundance is commonly ascribed to shock chemistry \citep{Schilke1997}.

\subsection{Limits on HCl absorption in the diffuse material}
The HCl column density, $N({\rm HCl})$,  we derive for the gas associated 
with Sgr~B2(M) (see Table \ref{t:hclmodel}) is in good agreement, i.e. 16\%\ higher, 
with the value derived by \citet{Zmuidzinas1995}. 

For the velocity components corresponding to the diffuse lines of sight 
we derive an upper limit on the HCl column density, $N({\rm HCl})$, 
of $8.4 \times 10^{12}$ cm$^{-2}$ (see Table~\ref{t:hclmodel} and 
Fig.~\ref{f:hcl} in \S\ref{sss:hcl}). 
Since $N({\rm H})$ ranges from $0.5$--$2.4 \times 10^{22}$~cm$^{-2}$
we calculate an HCl to hydrogen nuclei abundance ratio [HCl/H] of 
$0.6 - 3 \times 10^{-10}$. 
Our upper limit is much coarser than the $(2.7\pm1.0) \times 10^{11}$~cm$^{-2}$ 
\citet{Federman1995} determine for $N({\rm HCl})$ from an UV absorption line 
toward $\zeta$~Oph. 

\section{Conclusions}
We have reported submillimeter wavelength absorption from rotational ground-state 
transitions of $^{13}$CH$^+$, H$^{35}$Cl, H$^{37}$Cl, and, for the first time, 
of the SH$^+$ molecule toward the strong Galactic center continuum source Sgr~B2(M). 
From $^{13}$CH$^+$ and SH$^+$ we detect absorption at the systemic LSR velocity 
of Sgr~B2(M) and in addition over a wide range of lower velocities 
belonging to intervening material. 
For the intervening clouds we have derived molecular hydrogen column densities from 
optically thin H$^{13}$CO$^{13}$ emission and atomic column densities from HI 21 cm line data. 
For all the clouds we find that the former is larger than the latter by factors between 15 and 130.
If we ``scale'' the observed velocity ranges of the individual clouds to the line widths 
of single diffuse and translucent molecular clouds in the Solar neighborhood, we find similar $^{13}$CH$^+$ 
column densities in both types of clouds. Most likely the reason for this is that the observed $^{13}$CH$^+$ 
is indeed mostly associated with atomic material surrounding molecular material.
The [SH$^+$/CH$^+$] ratio is consistent with upper limits from optical data. 
Our upper limits on the HCl column density in the absorption components not associated 
with Sgr~B2 are consistent with but coarser than measurements of UV absorption from this molecule.

\begin{landscape}
 \begin{table}
 \caption{
 Column densities of the absorption components and abundances relative to H.
 }
\label{t:ratios}
 \begin{center}
 \begin{tabular}{cccccccccccccccccc}
 \hline\hline
 \noalign{\smallskip}
  & \multicolumn{2}{c}{SH$^+$} & & \multicolumn{2}{c}{CH$^+$} & & \multicolumn{2}{c}{SO} & & \multicolumn{2}{c}{H$_2$S} & & \multicolumn{2}{c}{SO$_2$} & & \multicolumn{2}{c}{CS}  \\ 
 \cline{2-3} \cline{5-6} \cline{8-9} \cline{11-12} \cline{14-15} \cline{17-18}
 \noalign{\smallskip}
 \multicolumn{1}{c}{Velocity range} & \multicolumn{1}{c}{$N$} & \multicolumn{1}{c}{$X_H$} & & \multicolumn{1}{c}{$N$} & \multicolumn{1}{c}{$X_H$} & & \multicolumn{1}{c}{$N$} & \multicolumn{1}{c}{$X_H$} & & \multicolumn{1}{c}{$N$} & \multicolumn{1}{c}{$X_H$} & & \multicolumn{1}{c}{$N$} & \multicolumn{1}{c}{$X_H$} & & \multicolumn{1}{c}{$N$} & \multicolumn{1}{c}{$X_H$} \\ 
  \multicolumn{1}{c}{\scriptsize (km~s$^{-1}$)} & \multicolumn{1}{c}{\scriptsize (cm$^{-2}$)} & & & \multicolumn{1}{c}{\scriptsize (cm$^{-2}$)} & & & \multicolumn{1}{c}{\scriptsize (cm$^{-2}$)} & & & \multicolumn{1}{c}{\scriptsize (cm$^{-2}$)} & & & \multicolumn{1}{c}{\scriptsize (cm$^{-2}$)} & & & \multicolumn{1}{c}{\scriptsize (cm$^{-2}$)} & \\ 
 \multicolumn{1}{c}{(1)} & \multicolumn{1}{c}{(2)} & \multicolumn{1}{c}{(3)} & & \multicolumn{1}{c}{(4)} & \multicolumn{1}{c}{(5)} & & \multicolumn{1}{c}{(6)} & \multicolumn{1}{c}{(7)} & & \multicolumn{1}{c}{(8)} & \multicolumn{1}{c}{(9)} & & \multicolumn{1}{c}{(10)} & \multicolumn{1}{c}{(11)} & & \multicolumn{1}{c}{(12)} & \multicolumn{1}{c}{(13)} \\ 
 \hline \\[-2.0ex]
 95.3 --  38.6& $  2.0\,(14)$ &   1.8(-10) & & $  3.6\,(14)$ &   3.3(-10) & & -- & -- & & -- & -- & & -- & -- & & -- & -- \\  
 23.2 --  12.9& $  3.2\,(13)$ &   2.9(-10) & & $  5.7\,(13)$ &   5.2(-10) & & -- & -- & & -- & -- & & -- & -- & & -- & -- \\  
-12.9 -- -33.5& $  3.4\,(13)$ &   1.2( -9) & & $  3.1\,(14)$ &   1.1( -8) &  & $  3.6\,(12)$ &   1.0(-10) &  & $  1.5\,(13)$ &   5.2(-10) &  & $ < 2.1\,(13)$ & $<$  7.3(-10) &  & $  1.6\,(13)$ &   5.5(-10) \\ 
-36.1 -- -51.5& $  1.5\,(13)$ &   3.9(-10) & & $  9.2\,(13)$ &   2.3( -9) &  & $  9.4\,(12)$ &   2.0(-10) &  & $  4.8\,(13)$ &   1.2( -9) &  & $ < 1.1\,(13)$ & $<$  2.7(-10) &  & $  2.8\,(13)$ &   7.1(-10) \\ 
-54.1 -- -64.4& -- & -- & & $  1.4\,(13)$ &   2.6( -9) & & -- & -- & & -- & -- & & -- & -- & & -- & -- \\  
-69.6 -- -85.0& $  3.9\,(13)$ &   2.7( -9) & & $  1.5\,(14)$ &   1.0( -8) &  & $ < 7.0\,(11)$ & $<$  5.0(-11) &  & $ < 4.2\,(12)$ & $<$  2.9(-10) &  & $ < 3.7\,(12)$ & $<$  2.6(-10) &  & $  1.3\,(13)$ &   9.1(-10) \\ 
-90.2 -- -123.7& $  6.9\,(13)$ &   2.2( -9) & & $  7.9\,(13)$ &   2.5( -9) &  & $  5.1\,(12)$ &   2.0(-10) &  & $ < 4.5\,(12)$ & $<$  1.4(-10) &  & $  1.9\,(13)$ &   6.1(-10) &  & $  4.4\,(13)$ &   1.4( -9) \\ 
 \hline
 \end{tabular}
 \end{center}
 Notes: $N$ is the column density of molecule Y and $X_H$ the column density ratio [Y]/([\ion{H}{i}]+$2\times$[H$_2$]). The column densities of SO, H$_2$S, and SO$_2$ are from Tieftrunk et al. (1994) and those of CS are from Greaves et al. (1992). The velocity ranges are the same as in Table~\ref{t:hi}. The list of absorption components used to compute the SH$^+$ and CH$^+$ column densities is given in Table~\ref{t:hi}. The components for which the column density is an upper limit in Table~\ref{t:shpmodel} were ignored. $x(y)$ means $x \times 10^y$.
 \end{table}
 \end{landscape}

\acknowledgements{We would like to thank the referee for a very through reading of the manuscripts and comments that led to significant improvements. H.S.P.M. is grateful to the Bundesministerium f\"ur Bildung
 und Forschung (BMBF) for financial support which was administered by the
 Deutsches Zentrum f\"ur Luft- und Raumfahrt (DLR). }

\bibliographystyle{aa}
\bibliography{KMM}

\clearpage

\end{document}